\documentclass[centertags,12pt]{amsart}
 
\usepackage{amssymb}
\usepackage{amsmath}
\usepackage{bm}
\usepackage{amsthm}
\usepackage[normalem]{ulem}
\usepackage{xcolor}
\colorlet{RED}{red}
\colorlet{BLUE}{blue}
\usepackage{graphicx}
\usepackage{comment}
\usepackage{lscape}
\usepackage{xcolor}
\definecolor{Cobalt}{rgb}{0.0, 0.28, 0.67}
 
\makeatletter
\renewcommand{\subsection}{\@startsection
{subsection}{2}{0mm}{\baselineskip}{-0.25cm}
{\normalfont\normalsize\em}}
\makeatother

\newtheorem{theorem}{Theorem}
\newtheorem{proposition}[theorem]{Proposition}
\newtheorem{corollary}[theorem]{Corollary}
\newtheorem{lemma}[theorem]{Lemma}
\theoremstyle{definition}
\newtheorem{definition}{Definition}
\newtheorem{example}{Example}
\theoremstyle{remark}
\newtheorem{remark}{Remark}

\newcommand{\ket}[1]{|#1 \rangle}
\newcommand{\bra}[1]{\langle #1 |}

\newcommand{\fp }{\mathbb{F}_p}
\newcommand{\fq }{\mathbb{F}_q}
\newcommand{\fqn }{\mathbb{F}_q^n}
\newcommand{\bba }{\bm{a}}
\newcommand{\bbb }{\bm{b}}

\newcommand{\bbv }{\bm{v}}
\newcommand{\bbw }{\bm{w}}

\newcommand{\bbz }{\bm{z}}

\newcommand{\cala }{\mathcal{A}}
\newcommand{\calc }{\mathcal{C}}
\newcommand{\cald }{\mathcal{D}}

\newcommand{\calq }{\mathcal{Q}}

\newcommand{\calr }{\mathcal{R}}

\title[Optimal pure quantum LRCs  from matrix-product construction]{Optimal pure quantum $(\MakeLowercase{r},\delta)$-locally recoverable codes  from matrix-product  construction}
\author[C. Galindo]{Carlos Galindo}
\address{Universitat Jaume I, Campus de Riu Sec, Departamento de Matem\'aticas \& Institut Universitari de Matem\`atiques i Aplicacions de Castell\'o, 12071
Caste\-ll\'on de la Plana, Spain.}\email{galindo@uji.es}

\author[F. Hernando]{Fernando Hernando}
\address{Universitat Jaume I, Campus de Riu Sec, Departamento de Matem\'aticas \& Institut Universitari de Matem\`atiques i Aplicacions de Castell\'o, 12071
Caste\-ll\'on de la Plana, Spain.}\email{carrillf@uji.es}

\author[C. Munuera]{Carlos Munuera}
\address{Universidad de Valladolid, IMUVA-Mathematics Research Institute, 47011 Valladolid, Spain.} \email{cmunuera@uva.es}

\author[ D. Ruano]{Diego Ruano}
\address{Universidad de Valladolid, IMUVA-Mathematics Research Institute, 47011 Valladolid, Spain.} \email{diego.ruano@uva.es}

\thanks{This work was partially supported by Grants PID2022-138906NB-C21 and PID2022-138906NB-C22 funded by MCIN/AEI/ 10.13039/501100011033 and  by ERDF/UE, and by grant GACUJIMA-2024-03 funded by Universitat Jaume I}


\begin{document}

\begin{abstract}
Locally recoverable codes (LRCs) are classical error-correcting codes widely used in large-scale distributed and cloud storage systems. Quantum locally recoverable codes of locality $(r,\delta)$ (quantum $(r,\delta)$-LRCs) are the quantum counterpart of classical $(r,\delta)$-LRCs. They allow us to correct erasures at several positions using a trace-preserving quantum operation acting on qudits of a larger set of positions. Quantum  $(r,\delta)$-LRCs, $\mathcal{Q}(\mathcal{C})$, can be constructed from classical Euclidean (or Hermitian) dual-containing codes $\mathcal{C}$, and their recovery abilities are upper bounded by the minimum distance of the Euclidean (or Hermitian) dual of those codes. Parameters and localities of pure quantum $(r,\delta)$-LRCs satisfy a Singleton-like bound;  codes attaining equality are referred to as optimal.

We consider matrix-product codes (MPCs)  $\mathcal{C}$ and give constituent (or defining) matrices and conditions on the constituent codes such that the codes  $\mathcal{C}$ satisfy the conditions to provide quantum $(r,\delta)$-LRCs. As a consequence, we are able to determine their locality and parameters. Furthermore, we determine families of  optimal pure quantum $(r,\delta)$-LRCs derived from them.
\end{abstract}

\maketitle

\section{Introduction}
\label{sect1}

Quantum computers are set to change the world. Nowadays there are already quite advanced quantum devices supported by several big companies. Quantum computers manage quantum information, and protecting it from quantum noise (such as decoherence) is an essential task. Quantum error-correcting codes (QECCs) are designed  to perform it.

Binary QECCs were first introduced  \cite{Calder1,Calder2,Gott}. Subsequently $q$-ary QECCs were brought in and studied for non-binary values of $q$. The literature on these codes is huge, some articles in this line are \cite{Aly, AND, Ashi, HER-HE,KKK,LAG}. The integer $q$ is a power of a prime $p$ and a $q$-ary QECC $\mathcal{Q}$ is simply a linear subspace of the complex Hilbert space $\mathbb{C}^{q^n} :=  \mathbb{C}^{q} \otimes \cdots \otimes \mathbb{C}^{q}$. The positive integer $n$ is the length of $\mathcal{Q}$. In this paper we are concerned with one of the most used families of QECCs which is that of {\it quantum stabilizer codes}. These codes consider an error group and are nonzero vector subspaces of $\mathbb{C}^{q^n}$ formed by the intersection of the eigenspaces with eigenvalue $1$ corresponding to a subgroup of the error group.

This family has the advantage that quantum codes are constructed from classical ones. In fact, the existence of a $q$-ary stabilizer code $\mathcal{Q}$ is equivalent to the existence of an additive  code $\mathcal{C}$ included in $\fq^{2n}$ which is self-orthogonal with respect to a trace-symplectic form on that linear space. The dimension of $\mathcal{Q}$ and its correction capability depend only on  $\mathcal{C}$ and its symplectic-dual \cite[Theorem 13]{KKK}. Particular cases of quantum stabilizer codes can be obtained by considering self-orthogonality with respect to the Euclidean inner product of vectors in $\fq^{n}$ or to the
Hermitian inner product of vectors in $\mathbb{F}_{q^2}^n$. We give some details in the first part of Subsection \ref{QRDLRC}.

Regarding classical codes, in 2001 Blackmore and Norton \cite{BlacNor} introduced a new technique to obtain large codes from a matrix and a family of codes, usually named constituent matrix and codes. The obtained codes are called {\it matrix-product} (MPCs) and some articles where they are studied are \cite{FAN, FLL, HLR, HER, ROD, SOB,VAN}. It is natural to use these codes to construct QECCs with large lengths and good parameters. This approach was initiated in \cite{GHR} and continued with quite a few papers, some of them are \cite{Cao2, Cao3, Cao1,LIUDIN} and the very recent \cite{Wang}.
Subsection \ref{matrixpc} defines MPCs and summarizes some known facts on the parameters of MPCs. In the course of the paper, we will use these facts.

{\it Locally recoverable codes} (LRCs) were introduced in \cite{GHSY}, motivated by the use of coding theory techniques applied to distributed and cloud storage systems where the information is disseminated among several nodes. The growth of the amount of stored data makes the loss of information due to node failures a major problem. To obtain a reliable storage, when a node fails, we should be able to recover the data by using information from the other nodes, this is the {\em repair problem}. This problem can be stated, in terms of coding theory, as follows: Let $\mathcal{C}$ be an $[n,k,d]_q$ linear code over a finite field $\fq$. A coordinate $i\in\{1,\dots,n\}$ is locally recoverable with locality $r$ if there is a recovery set $R\subset\{1,\dots,n\}$ of cardinality $r$, with $i\notin R$,  such that an erasure in position $i$ of any codeword $\bm{c}\in\mathcal{C}$ can be recovered by using the information given by the coordinates of $\bm{c}$ with indices in $R$. A {\em recovery structure} is a collection of recovery sets for all coordinates. A code $\mathcal{C}$ is an LRC with locality $r$ if any coordinate is locally recoverable with locality at most $r$. Usually, we will assume that $r$ is as small as possible in such a way that the above holds.

In the literature we can find several procedures to obtain long-length LRCs, see for instance \cite{CMST, Jin,KWG,LMC,LMD,LXY,Mi,SVAV,ZXL}. LRCs with a blockwise recovery structure are particularly interesting. In these codes, the set of coordinates is partitioned into disjoint blocks, $\{1,\dots,n\}=\overline{R}_1\cup\dots\cup \overline{R}_t$, in such a way that for any coordinate $j\in\overline{R}_i$, the set $\overline{R}_i\setminus\{j\}$ is a recovery set for $j$. Good examples of this type of LRCs are presented in \cite{TB,TPD}.

LRCs described above cannot cope with a failure at several nodes. There are several approaches to deal with more than one erasure \cite{handbook}. A popular one in the literature consists of considering $(r, \delta)$-locality. We say that a code $C$ is an $(r,\delta)$-LRC if for each code coordinate,  there is a punctured subcode of $C$ containing that coordinate with length $r+\delta -1$, dimension $r$ and minimum distance $\delta$. That is, the LRC can cope with $\delta -1$  erasures. In the case  $\delta=2$, we have a classical LRC. Some references for this type of LRCs are \cite{CXHF,FF,HE,GHM,PKLK,PLK,WZ,Zh}. A very succinct introduction to these codes is given in Subsection \ref{sscrc}. The locality and parameters of these codes satisfy a Singleton-like bound (see the forthcoming Inequality  (\ref{deltaSingletonEquation})) and LRCs attaining this bound are called optimal; a discussion on the known parameters and locality of codes of this type can be consulted in \cite{CAIFAN}. Bounds of Gilbert-Varshamov type are also given for LRCs, see \cite{ZHANG} and references therein.
Good LRCs with $(r,\delta)$-locality have been also constructed from MPCs \cite{LEL}. In this case, the constituent codes are nested and the attached matrix must be nonsingular by columns, a concept which we have recalled in Subsection \ref{matrixpc}.

Currently LRCs are classical error-correcting codes widely used in large-scale distributed and cloud storage systems. Since quantum computers are expected to be fully available in the near future, {\it quantum locally recoverable codes} (QLRCs) have been recently introduced. This first occurred for quantum $r$-LRCs \cite{GUR} (see also \cite{GUR2, SHAR}) and afterwards for quantum $(r,\delta)$-LRCs \cite{QLRC}.

One of the primary results in \cite{QLRC} states that a $q$-ary quantum $(r,\delta)$-LRC ($(r,\delta)$-QLRC
for short) can be constructed from a classical $q$-ary  (respectively, from a classical $q^2$-ary) $(r,\delta)$-LRC which is
Euclidean (respectively, Hermitian) dual-containing. Informally, a quantum code is called pure if its minimum distance coincides with that of the underlying classical code. Thus, when the code $\calq(\calc)$ is pure, its parameters can be derived from those of $\calc$. These pure quantum $(r,\delta)$-LRCs satisfy a Singleton-like inequality. Note that if $\calq(\calc)$ is constructed from a Euclidean dual-containing code $\calc$, the condition $d(\calc) < d(\calc^{\perp_E})$ implies the purity of $\calq(\calc)$; similarly if $\calc$ is Hermitian dual-containing, the condition $d(\calc) < d(\calc^{\perp_H})$ ensures purity.

As with classical $(r,\delta)$-LRCs, we say that $\calq(\calc)$ is {\it optimal} if it attains the aforementioned inequality. Further details are provided in Subsection \ref{QRDLRC}, specifically in the forthcoming Theorem \ref{Te:QLRC} and Definition \ref{44}. An important corollary of this construction is that if $\calc$  determines a pure $(r,\delta)$-QLRC and is itself an optimal classical $(r,\delta)$-LRC, then  $\calq(\calc)$ is an optimal pure quantum $(r,\delta)$-LRC.

In this paper we focus on pure quantum $(r,\delta)$-LRCs. Regarding quantum $(r,\delta)$-LRCs, the only known Singleton-like bound is the one provided in \cite{QLRC} for pure codes (see Theorem \ref{Te:QLRC}). For QLRCs of locality $r$ (the case when a single erasure can be corrected), a Singleton-like bound for general quantum codes is presented in  \cite[Theorem 35]{GUR2}. Additionally,  \cite[Corollary 5]{LUO} establishes a bound for CSS codes \cite{Calder1, STE} of locality $r$ which are not required to be pure.

The main goal of this article is to provide new optimal pure quantum $(r,\delta)$-LRCs. Our codes $\calq(\calc)$ use classical $(r,\delta)$-LRCs $\calc$ which are MPCs. Then, with the notation in  Subsection \ref{matrixpc}, $\mathcal{C} = \left[\mathcal{C}_1, \ldots,  \mathcal{C}_s\right] \cdot A$. The construction in \cite{LEL}, where the $(r,\delta)$-recovery structure is given by that of a linear code $\mathcal{D}$ such that $\calc_1,\dots,\calc_s\subseteq\mathcal{D}$, will be useful for treating our matrix-product codes as $(r,\delta)$-LRCs, see Subsection \ref{(r,delta)separados}. However we introduce a new tool, as the matrix $A$ itself can also serve as an $(r,\delta)$-recovery structure for matrix-product codes. This is explained in Subsection \ref{se:SeveralErasuresMB} and will be used in some of the families of $(r,\delta)$-QLRCs provided in the last section of the paper.

As indicated, we look for matrix-product codes satisfying the conditions in Theorem \ref{Te:QLRC}. Therefore,
we are interested in dual containment of MPCs. We divide our study into two cases: first, when the constituent matrix $A$ of the MPC is square; and second, the general case where no restrictions are imposed on the size of $A$.

The study of the second case can be regarded as an extension of the first one.  However, we prefer this treatment because it facilitates  the understanding of the second one and principally because the literature studied initially the square case giving rise to some open problems \cite{Cao1, Cao2, Cao3}. The square case is treated in our Subsection \ref{La331}, where we give a partial solution to the open Problems 5.2(a) and 5.2(b) in \cite{Cao2}; in fact we solve them for matrices of the largest possible size.

Theorems \ref{te: ConditionsEuclideanDualContaining}, \ref{te: ConditionsEuclideanDualContaining_1} and \ref{te: ConditionsHermitianDualContaining} within Subsection \ref{La332} explain how to get Euclidean and Hermitian dual-containing matrix-product codes and compute their parameters. Therefore, these codes are suitable to produce QECCs. A useful component for them are the results in Subsection \ref{La31} which mostly come from \cite{Dual MPC}.

The main contributions of the paper appear in Section \ref{La42}. Here one can find several results giving the parameters of quantum $(r,\delta)$-LRCs determined by suitable MPCs, where we consider both Euclidean and Hermitian inner products. On the one hand, Theorems \ref{te: MainEuclideanCase} and \ref{te: MainEuclideanCase2} —supported by Theorem \ref{te: ConditionsEuclideanDualContaining}— utilize Euclidean dual containment to determine pure quantum codes. On the other hand, Theorem \ref{te: MainHermitianCase} considers Hermitian duality, while Proposition \ref{LAA43} uses a different recovery procedure. Building on Theorem \ref{te: MainEuclideanCase}, Corollaries \ref{El36-3} and \ref{El36-4} provide optimal pure quantum $(r,\delta)$-LRCs with lengths that are perfect squares. Furthermore, by considering suitable integers $t$, which, along with the field size or a positive integer $h$, determine the $(r,\delta)$-locality, Theorems \ref{te: EuclideanOtimalQLRC} and \ref{EEl41} establish the parameters for numerous optimal pure quantum $(r,\delta)$-LRCs. QLRCs of this latter type, obtained via Hermitian dual containment, are provided in Corollary \ref{El46}. Table \ref{Tabla1} summarizes the parameters and localities of the QLRCs described in Section \ref{La42}.

To date, the literature on optimal pure quantum $(r,\delta)$-LRCs remains sparse. Apart from \cite{QLRC}, additional families were only recently obtained in \cite{ZHOU,CaoL}. Our optimal codes feature different localities compared to those in \cite{QLRC} and \cite{ZHOU,CaoL}, and are therefore novel. In addition, their parameters are highly versatile. As an illustration of our results, Subsection \ref{laultima} includes three tables, Tables \ref{Tabla2}, \ref{Tabla3} and \ref{Tabla4} featuring several pure quantum $(r,\delta)$-LRCs. As these tables indicate, Theorem \ref{EEl41} leverages the constituent matrix of many specific MPCs for recovery purposes; this approach is particularly fruitful, as it allows for the construction of many optimal pure codes.

In summary, we introduce new local recovery techniques for matrix-product codes (MPCs), including recovery properties derived from the constituent matrix, which we then translate into the context of quantum local recovery. Our results demonstrate that these techniques yield numerous new optimal pure quantum $(r,\delta)$-LRCs. It is worth noting that, at present, only pure quantum codes can be considered optimal for multiple local recovery.
 
\section{Background}
\label{sect2}

The main goal of this paper is to obtain optimal pure quantum $(r,\delta)$-locally recoverable codes, $(r,\delta)$-LRCs, from classical matrix-product codes (MPCs). In this section, we recall some definitions and facts concerning these concepts. Quantum $(r,\delta)$-LRCs are closely related to classical $(r,\delta)$-LRCs. For this reason, it is convenient to start with a brief introduction to the latter family of codes. Throughout this article, $\mbox{\rm wt}(\bbv)$ denotes the Hamming weight of a vector $\bbv$. 

\subsection{Classical locally recoverable codes}
\label{sscrc}

We mostly follow the usual conventions and definitions for locally recoverable codes (LRCs). Let $n$ and $i$ be positive integers such that $i \leq n$. Given a non-empty set $R\subset \{1,\dots,n\}$ such that $i\notin R$, we write $\overline{R}(i) = R\cup\{i\}$ (or simply $\overline{R}$ when $i$ is clear from the context). Consider an $[n, k, d]_q$ code $\mathcal{C}$  over a finite field $\mathbb{F}_q$ and a generator matrix $G$ of $\mathcal{C}$ whose columns are $\bm{g}_1,\dots,\bm{g}_n$. As our purpose in this article is local recovery,  we only consider non-degenerate codes, that is, codes such that  $\bm{g}_i\neq\bm{0}$ for all $i=1,\dots,n$.  The set $R$ is a {\em recovery set} for the coordinate $i\notin R$ if $\bm{g_i}$ is a linear combination of  $\{\bm{g_t} : t\in R\}$, or equivalently if $\mbox{rank}(G_R)=\mbox{rank}(G_{\overline{R}})$, where $G_R$ denotes the submatrix of $G$ whose columns are those with indices in $R$. In this case, for any codeword $\bm{c}=(c_1,\dots,c_n)\in\mathcal{C}$, the coordinate $c_i$ can be obtained from the coordinates $c_t$, $t\in R$.

Let $r$ be the cardinality of the set $R$ and let $\pi_R : \mathbb{F}_q^n \rightarrow \mathbb{F}_q^r$ be the projection on the coordinates in $R$.  Denote by $\mathcal{C}[R]$ the punctured  code $\{ \pi_R(\bm{c}) : \bm{c}\in \mathcal{C}\}$. Set $\dim (\mathcal{C})$ for the dimension of the code $\mathcal{C}$.  Clearly $R$ is a  recovery set for the coordinate $i\notin R$, if and only if $\dim(\mathcal{C}[R]) = \dim(\mathcal{C}[\overline{R}])$. Thus the notion of recovery set does not depend on the generator matrix chosen.

The smallest cardinality of a recovery set  for a coordinate $i$ is the {\em locality of $i$}. The {\em locality of $\mathcal{C}$}, often denoted by $r=r(\mathcal{C})$, is the largest locality of its coordinates. The following Singleton-like inequality can be found in \cite[Theorem 5]{GHSY}.

\begin{theorem}\label{d1Singleton}
The locality $r$ of an LRC  with parameters $[n, k, d]_q$ satisfies
\begin{equation}\label{d1SingletonEquation}
d+k+\left\lceil \frac{k}{r} \right\rceil \le n+2.
\end{equation}
\end{theorem}

The difference between the right- and the left-hand sides in (\ref{d1SingletonEquation}), $\Delta=n+2-d-k-\lceil k/r \rceil$,  is called the LRC-{\em Singleton defect} of $\mathcal{C}$. Codes with $\Delta = 0$ are called {\em Singleton-optimal} (or simply {\em optimal}).

Many good LRCs have a blockwise structure: the entire set of coordinate indices $\{1,\dots,n\}$ is partitioned into disjoint sets $\overline{R}$ that are {\em extended recovery sets}, so that for any coordinate $i\in\overline{R}$, the set $\overline{R}\setminus\{ i\}$ is a recovery set for $i$. This is the case of the well-known Tamo-Barg construction \cite{TB}. As we will see, matrix-product codes have, in a natural way, two simultaneous structures of this type.

The class of LRCs we have described allows for the local recovery of the information stored in the event of failure of a single node. However, concurrent failures of several nodes in a network are also possible and even common. Such failures have been handled in different ways in the literature. Here we consider the approach in \cite{PKLK} (see also
\cite{CH,CXHF, FF,GHM,PLK,WZ, Zh,ZL}. We say that an
 LRC  $\mathcal{C}$ has {\em locality $(r,\delta)$} (or that it is an $(r,\delta)$-LRC), if, for every coordinate $i$, there exists a set of positions  $\overline{R}=\overline{R}(i)\subseteq\{1,2,\dots,n\}$ containing $i$ such that \newline
\rule{10mm}{0mm} (RD1) $|\overline{R}|\le r+\delta-1$; and \newline
\rule{10mm}{0mm} (RD2) $d(\mathcal{C}[\overline{R}])\ge \delta,$

where $|\overline{R}|$ means cardinality of $\overline{R}$ and $d(\mathcal{C})$ denotes the minimum distance of the code $\mathcal{C}$.

A set $\overline{R}=\overline{R}(i)$  satisfying the above conditions (RD1) and (RD2) is  called an $(r,\delta)$-{\em extended recovery set} for the coordinate $i$.
Given such a set $\overline{R}$, the correction capability of $\mathcal{C}[\overline{R}]$ can be used to correct an erasure at any position $j\in\overline{R}$ in the presence of up to $\delta-2$ additional erasures in $\overline{R}$ (which are also recovered).  Notice that the classical definition of locality corresponds to the case $\delta=2$. Provided that $\delta\ge 2$, any subset $R\subseteq\overline{R}$ of cardinality $r$ with $i\notin R$ is a recovery set for $i$. Thus if $\mathcal{C}$ has locality $(r,\delta)$, then the classical locality of $\mathcal{C}$ is at most $r$.

The $(r,\delta)$-locality  satisfies the following generalization of the Singleton-like bound given in Theorem \ref{d1Singleton}, which was proved in \cite[Theorem 2]{PKLK}.

\begin{theorem}\label{drSingleton}
Let $\mathcal{C}$ be an LRC with parameters $[n, k, d]_q$ and locality $(r,\delta)$. Then, the following inequality holds
\begin{equation}\label{deltaSingletonEquation}
d+k+\left( \left\lceil \frac{k}{r}\right\rceil-1 \right) (\delta-1) \le n+1.
\end{equation}
\end{theorem}

Codes attaining the bound in $(\ref{deltaSingletonEquation})$ are called  $(r,\delta)$-{\em optimal}.

\subsection{Quantum $(r,\delta)$-locally recoverable codes}
\label{QRDLRC}

As above, let $q$ be a power of a prime $p$. A {\it $q$-ary quantum error-correcting code} (QECC) of length $n$ is a linear subspace of $\mathbb{C}^{q^n}$, where $\mathbb{C}$ denotes the field of complex numbers. In this paper, we are mainly interested in a class of quantum codes, called {\it quantum stabilizer codes.} These codes can be regarded as the intersection of the eigenspaces with eigenvalue $1$ determined by a commutative subgroup $S$ of the error group $G_n$ generated  by the nice error basis $\epsilon_n := \{ E_{(\bba,\bbb)}\; :\; \bba,\bbb \in \mathbb{F}_q^{n}\}$. $E_{(\bba,\bbb)}$ are error operators of the form  $X(\bba) Z(\bbb)$, where  $X(\bba)$ and $Z(\bbb)$ are given by tensoring $n$ unitary operators on $\mathbb{C}^{q}$ defined by $X(a)\ket{x} = \ket{x+a}$ and $Z(b)\ket{x}=\xi^{\mathrm{tr}(bx)}\ket{x}$. Here $\xi = e^{\frac{\iota 2 \pi}{p}}$, $\iota = \sqrt{-1}$, is a primitive $p$th root of unity; $\mathrm{tr}$ is the trace map from $\mathbb{F}_q$ to $\mathbb{F}_p$; $x, a, b \in \mathbb{F}_q$;  and the set $\{\ket{x}\}_{x \in \mathbb{F}_q}$ represents a distinguished orthonormal basis of $\mathbb{C}^{q}$, see \cite{KKK}. Thus, the {\it quantum stabilizer code} given by $S$ is $\calq(S) = \bigcap_{E \in S} \{ \upsilon \in \mathbb{C}^{q^n}\; :\; E \upsilon = \upsilon\}$.

Let $\bba,\bbb\in\fqn$,  $\bba = (a_1, \ldots, a_n)$ and $\bbb = (b_1, \ldots, b_n)$. The symplectic weight of the vector $(\bba|\bbb) \in \mathbb{F}_q^{2n}$ is defined as
$$
\mathrm{swt} (\bba|\bbb) := | \{ j : (a_j, b_j) \neq (0,0)\}|.
$$
In this context, the weight $\mathrm{wt}(E)$ of an operator $E = \xi^\ell E_{(\bba,\bbb)} \in G_n$ is given by the symplectic weight $\mathrm{swt}(\bba|\bbb)$. A quantum code $\calq$ is said to have minimum distance $d$ if it can detect all errors in
$G_n$ with weight less than $d$, but fails to detect at least one error of weight exactly $d$. A quantum stabilizer code of type $((n,K,d))_q$ is a $K$-dimensional subspace of $\mathbb{C}^{q^n} = \mathbb{C}^q \otimes \cdots \otimes \mathbb{C}^q$ with minimum distance $d$. When $K = q^k$, such a code is said to be an $[[n,k,d]]_q$ stabilizer code. Moreover, a quantum stabilizer code is {\it pure} if its stabilizer $S$ does not contain non-scalar matrices of weight less than $d$.

Recall that for vectors $\bba, \bbb \in \fqn$, the {\it Euclidean inner product} is defined as
$$
\langle \bba, \bbb \rangle_E := \sum_{j=1}^{n} a_j b_j.
$$
If $\bba, \bbb \in \mathbb{F}_{q^2}^{n}$,  the {\em Hermitian inner product} is given by
$$
\langle \bba, \bbb \rangle_H  := \sum_{j=1}^{n} a_j b_j^q.
$$

For a linear code $\calc$ over $\mathbb{F}_{q}$, we denote by $\calc^{\perp_E}$  the Euclidean dual of $\calc$ with respect to the inner product $\langle \cdot \rangle_E$. If  $\calc$ is defined over $\mathbb{F}_{q^2}$, then $\calc^{\perp_H}$ denotes its Hermitian dual code. $\calc$ is said to be {\em dual-containing} if $\calc^{\perp_E} \subseteq \calc$, and {\em Hermitian dual-containing} if $\calc^{\perp_H} \subseteq \calc$. Let us state two well-known results that show how to find quantum stabilizer codes from classical linear ones \cite[Corollary 19 and Lemma 20]{KKK}, see also \cite{Calder1, STE}.

\begin{theorem}
\label{resto}
The following facts hold:
\begin{enumerate}
\item The existence of an  $[n,k,d]_{q^2}$ Hermitian dual-containing linear code $\calc$,  implies that of an $[[n, 2k-n, \geq d]]_q$ quantum stabilizer  code, $\calq(\calc)$.
\item The existence of two $q$-ary linear codes $\calc_j$ with parameters $[n,k_j,d_j]_q$, $j=1,2$, such that $\calc_2^{\perp_E} \subseteq \calc_1$ implies the existence of an $[[n, k_1+k_2-n,d]]_q$ quantum stabilizer code $\calq(\calc_1,\calc_2)$, with $d= \min \{\mathrm{wt}(\bbz) : \bbz \in (\calc_1 \setminus \calc_2^{\perp_E}) \cup (\calc_2 \setminus \calc_1^{\perp_E})\}$.
\end{enumerate}
\end{theorem}

When $\calc_1=\calc_2=\calc$, we simply write  $\calq(\calc)$ instead of $\calq(\calc_1,\calc_2)$.

We now formalize when a set of erasures can be corrected by a quantum stabilizer code $\calq$. For a complex matrix  $A$, we denote by $A^\dagger$ its conjugate transpose (also known as the Hermitian adjoint).
For any $q \times q$ complex matrix $\rho$, define the \emph{completely depolarizing channel} on a single qudit in $\mathbb{C}^q$ as follows:
\[
\Gamma(\rho) = \frac{1}{q^2} \sum_{E \in \epsilon_1} E \rho E^\dagger,
\]
where $\epsilon_1$ is the previously introduced nice error basis acting on $\mathbb{C}^q$. This channel models a scenario in which each error operator in $\epsilon_1$, including the identity matrix ${\rm I}_q$, is applied with equal probability $1/q^2$. Now consider a quantum code $\calq \subseteq \mathbb{C}^{q^n} = \mathbb{C}^q \otimes \cdots \otimes \mathbb{C}^q$, and let $I \subseteq \{1, \ldots, n\}$ be a subset of qudit positions. Define the map $\Gamma^I = \bigotimes_{j=1}^n \Gamma_j$, where each $\Gamma_j$ is given by $\Gamma$ if $j \in I$, and by ${\rm I}_q$ otherwise;
in other words, $\Gamma$ is applied to qudits at the positions indexed by $I$.

By definition, \emph{erasures at the positions in $I$ are correctable by $\calq$} if there exists a recovery operation $\mathcal{R}_{Q,I}$ —a trace-preserving quantum operation as described in \cite[Chapter 8]{chuangnielsen}— such that for every $\ket{\varphi} \in \calq$, it holds that
$$
  \mathcal{R}_{Q,I} \circ \Gamma^I (\ket{\varphi}\bra{\varphi}) = \ket{\varphi}\bra{\varphi}. \label{eq1}
$$

Next, we recall the concept of quantum code $\calq \subseteq \mathbb{C}^{q^n}$  correcting  erasures at the positions in $I \subseteq \{1, \ldots, n\}$ using  the qudits in a set
$J \varsupsetneq I$.

\begin{definition}
\label{I-J}
Let $I,J$ be two set of indices such that
$\emptyset \neq I \varsubsetneq J \subseteq \{1, \ldots, n\}$. A quantum stabilizer code  $\calq \subseteq \mathbb{C}^{q^n}$ is an {\it $(I,J)$-locally recoverable code} if there exists a trace-preserving quantum operation $\mathcal{R}_{Q,I}^{J}$, which acts only on the qudits corresponding to $J$ and keeps untouched the remaining ones, such that
$$
  \mathcal{R}_{Q,I}^J \circ \Gamma^I (\ket{\varphi}\bra{\varphi}) = \ket{\varphi}\bra{\varphi} \label{eq2}
$$
for all $\ket{\varphi} \in \calq$.
\end{definition}

This definition allows us to introduce the concept of quantum $(r,\delta)$-locally recoverable code.

\begin{definition}
\label{lcr}
A quantum stabilizer code  $\calq \subseteq \mathbb{C}^{q^n}$ is called  a {\it quantum $(r,\delta)$-locally recoverable code} (quantum $(r,\delta)$-LRC or $(r,\delta)$-QLRC for short) if for each index $i \in \{1, \ldots, n\}$, there exists a set $J \subseteq \{1, \ldots, n\}$ containing $i$, with $|J| \leq r + \delta-1$, satisfying that for all subsets $I \subseteq J$ of cardinality $\delta -1$, the code $\calq$ is $(I,J)$-locally recoverable.
\end{definition}

The following result has been recently stated in \cite[Theorem 31]{QLRC}.

\begin{theorem}
\label{Te:QLRC}
Let $\calc$ be a $q$-ary  linear code included in $\mathbb{F}_{q}^n$ (respectively, a  $q^2$-ary linear code included in $\mathbb{F}_{q^2}^n$). Assume that $\calc$ is Euclidean (respectively, Hermitian) dual-containing, that $\dim (\calc) = \frac{n+k}{2}$, and  that $\calc$ is an $(r,\delta)$-LRC. Then, the quantum  code $\calq(\calc)$ is an $(r,\delta)$-QLRC with parameters $[[n,k, \geq d(\calc)]]_q$, satisfying the following inequality
\begin{equation}
\label{eq17}
k + 2 d(\calc) + 2\left(\left\lceil\frac{n+k}{2r}\right\rceil-1\right)(\delta-1) \leq n+2.
\end{equation}
\end{theorem}
 
In both the Euclidean and the Hermitian cases, $d(\calq(\calc)) = \mathrm{wt} (C \setminus C^\perp)$, where $\perp$ denotes the Euclidean or Hermitian dual depending on the case considered, and $\mathrm{wt} (C \setminus C^\perp)$ is the minimum Hamming weight of the non-vanishing elements in $C \setminus C^\perp$. The code $\calq(\calc)$ is pure if and only if $d(\calq(\calc)) = d(\calc)$, and, in this case, the Inequality (\ref{eq17}) can be interpreted in terms of the parameters of the quantum code $\calq(\calc)$.
This leads to the following definition given in \cite[Definition 33]{QLRC}.

\begin{definition}
\label{44}
A pure quantum $(r, \delta)$-LRC $\calq(\calc)$, as defined previously, is called {\it optimal} if its parameters and $(r, \delta)$-locality meet the bound given in Inequality~(\ref{eq17}).
\end{definition}

Note that our definition of optimality is with respect to the Singleton-like bound. It holds that if $\calc$ is an optimal $(r, \delta)$-LRC in the classical sense and gives rise to a pure quantum code $\calq(\calc)$, then $\calq(\calc)$ is an optimal pure quantum $(r, \delta)$-LRC \cite{QLRC}.

\subsection{Matrix-product codes}
\label{matrixpc}

We mainly follow the approach by Blackmore and Norton in \cite{BlacNor}, with a somewhat different notation.
Let $A=(a_{ij})$ be an $s\times h$ matrix with entries in $\fq$, where $2\le s\le h$. Given $\bm{v}_1,\dots,\bm{v}_s\in\fq^m$ with coordinates $\bm{v}_i=(v_{i1},\dots,v_{im})$, $i=1,\dots,s$, we consider the element $(\bm{v}_1,\dots,\bm{v}_s)\in(\fq^m)^s$ and define the product
$$
(\bm{v}_1,\dots,\bm{v}_s)\cdot A=\left( \sum_{i=1}^s a_{i1}\bm{v}_i, \dots, \sum_{i=1}^s a_{ih}\bm{v}_i  \right) \in(\fq^m)^h.
$$
For example,  the product
$$
(\bm{u},\bm{v}) \cdot
\left( \begin{array}{cc}
1 & 1 \\
0 & 1
\end{array} \right)
=(\bm{u},\bm{u}+\bm{v}),
\hspace*{7mm} \mbox{with~} \bm{u}, \bm{v}\in\fq^m,
$$
gives the well-known $(\bm{u},\bm{u}+\bm{v})$ Plotkin construction.

We write $\bm{v}=(\bm{v}_1,\dots,\bm{v}_s)$ and regard $\bm{v}$ as an element of $\fq^{ms}$. Each $\bm{v}_i$ is called a {\em block} of $\bm{v}$. Similarly, we write $\bm{v}\cdot A=(\bm{p}_1,\dots,\bm{p}_h)=\bm{p}$, with $\bm{p}_j=(p_{j1},\dots,p_{jm})$, $j=1,\dots,h$. The vector $\bm{p}$ is regarded as an element of $\fq^{mh}$ and each $\bm{p}_j$ is a {\em block} of $\bm{p}$.

Also, as a notation, for $\ell=1,\dots,m$, we write $\bm{v}^\ell=(v_{1\ell},\dots,v_{s\ell})\in\fq^s$, the vector consisting of the $\ell$th coordinates of $\bm{v}_1,\dots,\bm{v}_s$.  Similarly, $\bm{p}^\ell :=(p_{1\ell},\dots,p_{h\ell})\in\fq^h$.
The following property follows easily from the above definitions.

\begin{lemma}\label{lemmampc}
Keep the above notation and let $\bm{p}=\bm{v}\cdot A$. Then $\bm{p}^\ell=\bm{v}^\ell A$ for all $\ell=1,\dots,m$. Thus, if $A$ has full rank $s$, then $\bm{p}^\ell\neq \bm{0}$ whenever $\bm{v}^\ell\neq \bm{0}$.
\end{lemma}

As a consequence of Lemma \ref{lemmampc}, the map $\bm{v} \mapsto \bm{v}\cdot A$ is linear and injective when  $A$ has full rank $s$.

Let $\calc_1,\dots,\calc_s$ be $q$-ary non-degenerate linear codes of the same length $m$ and let $A$  be an $s\times h$ full rank matrix with entries in $\fq$.  The {\em matrix-product code} (MPC) $[\calc_1,\dots,\calc_s]\cdot A$ is defined as
$$
\calc=[\calc_1,\dots,\calc_s]\cdot A=\{ (\bm{v}_1,\dots,\bm{v}_s)\cdot A\, :\, \bm{v}_i\in \calc_i,~ i=1,\dots,s\}.
$$
We index the coordinates in $\mathcal{C}$ by using pairs $(j,\ell)$ with $1\le j\le h$ and $1\le \ell \le m$. Then, one gets the following result:

\begin{proposition}\label{mpcparametros}
With the above notation, $[\calc_1,\dots,\calc_s]\cdot A$ is a linear code of length $n=mh$,  dimension $k=\dim(\calc_1)+\dots+\dim(\calc_s)$, and generator matrix
\begin{equation*} 
G=\left(
\begin{array}{cccc}
a_{11}G_1 & a_{12}G_1 & \dots & a_{1h}G_1 \\
a_{21}G_2 & a_{22}G_2 & \dots & a_{2h}G_2 \\
\vdots & \vdots & & \vdots \\
a_{s1}G_s & a_{s2}G_s & \dots & a_{sh}G_s
\end{array} \right),
\end{equation*}
where $G_1,\dots,G_s$ are generator matrices of the codes $\calc_1,\dots,\calc_s$, respectively.
\end{proposition}

In order to determine the minimum distance of the code $[\calc_1,\dots,\calc_s]\cdot A$, we consider the linear codes $\cala_i$ generated by the $i$ first rows of $A$, $i=1\dots,s$. The following proposition collects results from \cite[Theorem 1]{HLR} and \cite{OS}. 

\begin{proposition}\label{pro:distance}
The minimum distance $d(\calc)$ of $\calc=[\calc_1,\dots,\calc_s]\cdot A$ satisfies the following inequality
$$
d(\calc) \ge \min \,\{ d(\calc_1)d(\cala_1),\dots, d(\calc_s) d(\cala_s)\},
$$
where, for $1 \leq i \leq s$, $d(\calc_i)$ (respectively, $d(\cala_i)$) denotes the minimum distance  of the code $\calc_i$ (respectively, $\cala_i$).

Furthermore, if the codes $\calc_i$ are nested, $\calc_1\supseteq\calc_2\supseteq\dots\supseteq \calc_s$, then equality holds.
\end{proposition}

The matrix $A$ is said to be {\em non-singular by columns} (NSC) whenever all the codes $\cala_i$, $1\le i\le s$, are MDS; that is, for all index $i$ and for all indices $1\le j_1<j_2<\cdots <j_i\le h$, the submatrix of $A$ obtained from the rows $1,\dots,i$, and the columns $ j_1,j_2,\ldots ,j_i$, is non-singular. In this case, $d(\cala_i)=h+1-i$ and then
$$
d( [\calc_1,\dots,\calc_s]\cdot A)\ge \min \{ hd(\calc_1),\dots, (h+1-s) d(\calc_s)\}.
$$

A simple way to construct NSC matrices is as follows. Consider  a set $\{\alpha_1,\dots,\alpha_h\}$ of distinct elements in $\fq$ and another integer $s$ such that $2\le s\le h\le q$. Then, the Vandermonde type matrix
\begin{equation}
\label{matrizRS}
A=\left(
\begin{array}{cccc}
1                    & 1                     & \cdots & 1 \\
\alpha_1         & \alpha_2          & \cdots & \alpha_h \\
\vdots             & \vdots              &           & \vdots \\
\alpha_1^{s-1} & \alpha_2^{s-1} & \cdots & \alpha_h^{s-1}
\end{array} \right)
\end{equation}
is NSC.  In the following we fix an enumeration $\{\alpha_1,\dots,\alpha_{q}\}$ of the elements in $\fq$,  and denote by $A(h,s)$ the above matrix (\ref{matrizRS}) obtained from the set $\{\alpha_1,\dots,\alpha_{h}\}$. Moreover, $\mathrm{RS}(h,s)$ stands for the Reed-Solomon code generated by the matrix $A(h,s)$. There are no NSC matrices with entries in $\fq$ of larger size, see \cite{BlacNor}.

\begin{proposition}
\label{prop6}
\cite[Proposition 3.3]{BlacNor} Let $s$ and $h$ be integers such that $2 \leq s \leq h$. Then, there exist an  NSC matrix with entries in $\fq$ of size $s\times h$ if and only if $h\le q$.
\end{proposition}

\section{$(r,\delta)$-LRCs from matrix-product codes}
\label{La4}

In this section we give three types of $(r,\delta)$-recovery structures for the matrix-product code  construction. Later, in the next section,  these results will be used  for obtaining quantum $(r,\delta)$-LRCs.
We use the notation introduced in Section \ref{sect2}.

\subsection{$(r,\delta)$-locality using the blocks separately}
\label{(r,delta)separados}

Let $\mathcal{C} = \left[\mathcal{C}_1, \ldots,  \mathcal{C}_s\right] \cdot A$ be an MPC. When  the component codes $\calc_i$  are subcodes of some linear code $\mathcal{D}$, then an $(r,\delta)$-recovery structure for $\mathcal{D}$ provides a recovery structure for $\calc$, with the same locality.

\begin{lemma}\label{recoveringseparately1}
Let $\mathcal{D}$ be a $q$-ary linear code such that $\calc_1,\dots,\calc_s\subseteq\mathcal{D}$.
For $j=1,\dots,h$, let $B_j=\{(j,1),\dots,(j,m)\}$ be the set of positions corresponding to the $j$th block.  Then, the punctured code   $\calc[B_j]$ satisfies $\calc[B_j]\subseteq \mathcal{D}$.
\end{lemma}
\begin{proof}
The statement follows from the fact that $\calc_1,\dots,\calc_s\subseteq\mathcal{D}$.
\end{proof}

\begin{proposition}\label{recoveringseparatelyrdelta}
Let $\mathcal{D}$ be a linear code such that $\calc_1,\dots,\calc_s\subseteq\mathcal{D}$.
If $\overline{R}\subseteq\{ 1,\dots,m\}$ is an $(r,\delta)$-extended recovery set for a coordinate $\ell$ of $\mathcal{D}$, then for $1\le j \le h$, the translate $\overline{R}_j=\{ (j,t) : t \in \overline{R}\}$ is an $(r,\delta)$-extended recovery set for the coordinate  $(j,\ell)$ belonging to the $j$th block of $\calc$.
Thus, if $\mathcal{D}$ is an LRC with locality $(r,\delta)$, then $\calc$ is an LRC with locality $(r,\delta)$.
\end{proposition}
\begin{proof}
According to Lemma \ref{recoveringseparately1}, we have that  $\calc[\overline{R}_j] =(\calc[B_j])[\overline{R}_j]\subseteq \mathcal{D}[\overline{R}]$.  Clearly the equality of cardinalities $| \overline{R}_j|=| \overline{R}|$ holds.  Since the codes $\calc_1,\dots,\calc_s$ are non-degenerate and $A$ has full rank, we get $\calc[\overline{R}_j]\neq \{\mathbf{0}\}$ and thus  $d(\calc[\overline{R}_j] )\geq d(\mathcal{D}[\overline{R}])$.
Then, the  conditions (RD1) and (RD2)  stated in Subsection \ref{sscrc}  are fulfilled, hence $\overline{R}_j$ is an $(r,\delta)$-extended recovery set for the coordinate  $(j,\ell)$.
\end{proof}

MPCs of this type were studied  in  \cite{LEL}, under the assumption that the matrix $A$ is nonsingular by columns and $\calc_1\supseteq\ \calc_2\supseteq\dots\supseteq \calc_s$. An analogous recovery procedure as in Proposition \ref{recoveringseparatelyrdelta}, is used in \cite{LEL}.

\subsection{$(r,\delta)$-locality with several blocks}
\label{se:SeveralErasuresMB}

Besides the previous one, MPCs admit a second recovery  structure arising from the matrix $A$ used to construct $\calc$.   If the code $\cala$ generated by the matrix $A$  has locality $(r,\delta)$, then its recovery structure can be translated to the matrix-product code  $\calc=[\calc_1,\dots,\calc_s]\cdot A$, as the next proposition shows.

\begin{proposition}
Let $\overline{R}\subseteq\{1,\dots,h\}$ be an $(r,\delta)$-extended recovery set for a coordinate
$j$, $1\le j \le h$, in the code $\cala$. Then, the set $\overline{R}^\ell=\{ (t,\ell) : t\in \overline{R}\}$ is an $(r,\delta)$-extended recovery set  for the coordinate $(j,\ell)$ in the code $\calc=[\calc_1,\dots,\calc_s]\cdot A$.
\end{proposition}
\begin{proof}
Let $B^\ell=\{ (j,\ell) : 1\le j \le h \}$. According to Lemma \ref{lemmampc}, it holds that $\bm{p}^\ell\in\cala$ for every $\bm{p}\in\mathcal{C}$. Thus $\calc[B^\ell]\subseteq\cala$ and $\calc[\overline{R}^\ell]=(\calc[B^\ell])[\overline{R}^\ell]\subseteq\cala[\overline{R}]$. Then the minimum distance of $\calc[\overline{R}^\ell]$ is at least $\delta$, and thus
$\overline{R}^\ell_j$ is an $(r,\delta)$-extended recovery set  for $(j,\ell)$ in the code $\calc$.
\end{proof}

\begin{corollary}
\label{coro12}
If the code $\cala$ generated by the matrix $A$ is an LRC with locality $(r,\delta)$,  then  the matrix-product code $\calc=[\calc_1,\dots,\calc_s]\cdot A$ also is an LRC with locality $(r,\delta)$.
\end{corollary}

\subsection{$(r,\delta)$-LRCs from variations of matrix-product codes}
\label{sect4}

In this subsection we provide $(r,\delta)$-LRCs from sove variations extending  MPCs.
For our purposes in this subsection, a linear $[n,k,d]_q$ code $\cald$ is called {\em enlargable}  if an extra last coordinate can be added to each codeword of $\cald$, so that the set $\hat{\cald}$ of all enlarged words forms a linear code with parameters $[n+1,k,d+1]_q$. For example, classical extended codes belong to this type when its minimum distance increases. Likewise, Reed-Solomon codes of length $n \le q$ can be enlarged by evaluating at one extra point.

Let $\calc_1$ and $\calc_2$, be two enlargable linear codes with parameters $[m,k_i,d_i]_q$, $i=1,2$, and enlargements $\hat{\calc_1},\hat{\calc_2}$. Let $G_1,G_2,\hat{G_1},\hat{G_2}$ be generator matrices of these codes. Given a  $2\times h$ NSC matrix $A=(a_{ij})$, we consider the code $\hat{\calc}$ having generator matrix
$$
\hat{G}=\left(
\begin{array}{cccc}
a_{11}\hat{G}_1 & a_{12}G_1 & \cdots & a_{1h}G_1 \\
a_{21}\hat{G}_2 & a_{22}G_2 & \cdots & a_{2h}G_2 \\
\end{array} \right).
$$

\begin{proposition}\label{prop:ext1}
Keeping the above notation, let $\calc_1$ and $\calc_2$ be two enlargable linear codes with parameters $[m,k_i,d_i]_q$, $i=1,2$, and enlargements $\hat{\calc_1}$ and $\hat{\calc_2}$, respectively. Assume that $\calc_1\supseteq \calc_2$ and $\hat{\calc_1}\supseteq \hat{\calc_2}$. Then the code $\hat{\calc}$ whose generator matrix is $\hat{G}$, has parameters
$$
[mh+1,k_1 + k_2, \ge\min\{d_1 h +1, d_2 (h-1) \}]_q.
$$
Moreover, if the code $\cala$ given by the matrix $A$ has locality $(r_\cala,\delta_\cala)$, and $\hat{\calc}_1$ has locality $(\hat{r}_1,\hat{\delta}_1)$, then the code $\hat{\calc}$ has locality $(r,\delta)$, where $r=\max\{r_\cala,\hat{r}_1\}$ and $\delta = \min \{ \delta_\cala, \hat{\delta}_1 \}$.
\end{proposition}
\begin{proof}
The statements about length and dimension of $\hat{\calc}$ are clear. With regard to the minimum distance, let $\boldsymbol{p}=(\boldsymbol{\hat{v}}_1,\boldsymbol{v}_2)\cdot A\in \hat{\calc}$ be a nonzero codeword. If $\boldsymbol{v}_2 = \boldsymbol{0}$, then $\boldsymbol{p}=(a_{11}\hat{\boldsymbol{v}}_1,a_{12} \boldsymbol{v}_1\dots,a_{1h}\boldsymbol{v}_1)$, where $\hat{\boldsymbol{v}}_1\in\calc_1$ is nonzero and $\boldsymbol{v}_1$ is obtained from $\hat{\boldsymbol{v}}_1$ by deleting its last coordinate. Then, recalling that $\cala_1$ (respectively, $\cala_2$)
is the code generated by the first row (respectively, the first two rows) of the matrix $A$ and since $d(\hat{\calc}_1) = d(\calc_1)$, one proves that $\mbox{wt}(\boldsymbol{p})\ge d_1d(\cala_1)+1$. Otherwise, $\hat{\calc}_1 \supseteq \hat{\calc}_2$ shows that one can get a vector $\boldsymbol{p}$ in $\hat{\calc}$ where the first block of coordinates vanishes and therefore $\mbox{wt}(\boldsymbol{p})\ge d_2d(\cala_2)$. Finally, $d(\cala_2) \geq h$ because $A$ is NSC.

To prove the last statement it suffices to notice that an extended recovery set for a coordinate $(j,\ell)$ is
a set of type  $\overline{R}_j$ given by the LRC structure of $\calc_1$ when $(j,\ell)= (1,m+1)$, as defined in Subsection \ref{(r,delta)separados}; and
a set of type $\overline{R}^\ell$ given by the LRC structure of the code $\cala$ when $(j,\ell)\neq (1,m+1)$, as seen in Subsection \ref{se:SeveralErasuresMB}.
\end{proof}

The same idea can be applied  $s$  ($\le h$) times. Given $\calc_1$, $\calc_2$ and $A$ as above, we can consider the code $\hat{\calc}$ with generator matrix:
$$
\hat{G}=\left(
\begin{array}{cccccc}
a_{11}\hat{G}_1 & \cdots & a_{1s}\hat{G}_1 &  a_{1{s+1}} G_1 & \cdots & a_{1h}G_1 \\
a_{21}\hat{G}_2 & \cdots & a_{2s}\hat{G}_2 &  a_{2{s+1}} G_2 & \cdots & a_{2h}G_2 \\
\end{array} \right),
$$
for which we have the following result, whose proof is similar to that of Proposition \ref{prop:ext1}.

\begin{proposition}\label{prop:ext2}
Keeping the above notation, let $\calc_1$ and $\calc_2$  be two enlargable linear codes with parameters $[m,k_i,d_i]_q$, $i=1,2$, and enlargements $\hat{\calc_1}$ and $\hat{\calc_2}$, respectively. Assume that $m<q$,  $\calc_1\supseteq \calc_2$ and $\hat{\calc_1}\supseteq \hat{\calc_2}$. Then the code $\hat{\calc}$  has parameters
$$
[mh+s,k_1 + k_2, \ge\min\{d_1 h +s, d_2 (h-1) + s-1 \}]_q.
$$
Moreover, if the code $\cala$ given by the matrix $A$ has locality $(r_\cala,\delta_\cala)$, and $\hat{\calc}_1$ has locality $(\hat{r}_1,\hat{\delta}_1)$, then the code $\hat{\calc}$ has locality $(r,\delta)$, where $r=\max\{r_\cala,\hat{r}_1\}$ and $\delta = \min \{ \delta_\cala, \hat{\delta}_1 \}$.
\end{proposition}

\section{Dual containment of MPCs}
\label{La3}

The main goal of this paper is to construct quantum $(r,\delta)$-LRCs from MPCs. Since
Euclidean and Hermitian dual containment conditions for MPCs are relevant for it, in this section we give some results supplying conditions of that type.

\subsection{Conditions for dual containment of MPCs}
\label{La31}

Let $\mathcal{C} = [\calc_1,\dots,\calc_s]\cdot A$ be an MPC. The dual code $\mathcal{C}^{\perp_E}$ was studied in \cite{BlacNor} when $A$ is a square matrix. Recently,  both $\mathcal{C}^{\perp_E}$ and $\mathcal{C}^{\perp_H}$ have been described for any matrix $A$. Consequently, conditions characterizing Euclidean and Hermitian dual containment of MPCs are also given. Propositions \ref{te: EuclideanDualMPC} and \ref{te: HermitianDualMPC}, and Theorems \ref{EuclideanDualContMPC} and \ref{HermitianDualContMP} recall some of these results which can be found in  \cite[Theorems 3 and 14]{Dual MPC}. 

Given a matrix $B$ with $t$ rows, and $ r \leq t$, we denote by $B\{1,\ldots,r\}$ the submatrix of $B$ consisting of its first $r$ rows and by  $B^{\tt t}$ the transpose of $B$.

\begin{proposition}
\label{te: EuclideanDualMPC}
Let $A$ be an $s \times h$ matrix of full rank $s$, with entries in $\mathbb{F}_{q}$, and let $\mathcal{C}_1,  \ldots,  \mathcal{C}_s$ be codes over $\fq$.
The  Euclidean dual of  the matrix-product code $\mathcal{C}=\left[\mathcal{C}_1,  \ldots,  \mathcal{C}_s\right] \cdot A$ is the matrix-product code
\begin{equation*}
\mathcal{C}^{\perp_E}= [\mathcal{C}^{\perp_E}_1,  \ldots, \mathcal{C}^{\perp_E}_s,  \overbrace{\mathcal{F},  \ldots,  \mathcal{F}}^{\text{$h-s$ times}} ] \cdot \left(B^{-1}\right)^{\tt t},
\end{equation*}
where $\mathcal{F} = \mathbb{F}_q^m$ and $B$ is any $h\times h$ invertible matrix over $\mathbb{F}_q$ satisfying $B\{1,\ldots,s\}=A$.
\end{proposition}

A similar result holds for Hermitian duality.

\begin{proposition}
\label{te: HermitianDualMPC}
Let $A$ be an $s \times h$ matrix of full  rank $s$, with entries in $\mathbb{F}_{q^2}$, and let $\mathcal{C}_1,  \ldots,  \mathcal{C}_s$ be codes over $\mathbb{F}_{q^2}$.
The  Hermitian dual of  the matrix-product code $\mathcal{C}=\left[\mathcal{C}_1,  \ldots,  \mathcal{C}_s\right] \cdot A$ is the matrix-product code
\begin{equation*}
\mathcal{C}^{\perp_H}= [\mathcal{C}_1^{\perp_H},  \ldots,  \mathcal{C}^{\perp_H}_s,  \overbrace{\mathcal{F},  \ldots.  \mathcal{F}}^{\text{$h-s$ times}} ] \cdot \left((B^{q})^{-1}\right)^{\tt t},
\end{equation*}
where $\mathcal{F} = \mathbb{F}_{q^2}^m$ and $B$ is any $h\times h$ invertible matrix over $\mathbb{F}_{q^2}$ satisfying $B\{1,\ldots,s\}=A$.
\end{proposition}

Next, let us state the conditions for dual containment of MPCs. First we describe the Euclidean case.

\begin{theorem}
\label{EuclideanDualContMPC}
Keeping the above notation,  let $\mathcal{C}=\left[\mathcal{C}_1,  \ldots, \mathcal{C}_s\right] \cdot A$, where $ A$ is of rank $s$ and it has entries in $\mathbb{F}_{q}$. Consider an $h \times h$ invertible matrix $B$ over $\mathbb{F}_{q}$ such that
$B\{1,\ldots,s\}=A$. Write $\left(B B^{\tt t}\right)^{-1}=\left(\zeta_{t,j}\right)_{1 \leq t, j \leq h}$. Then, $\mathcal{C}$ is Euclidean dual-containing if and only if the following conditions hold.
\begin{enumerate}
\item $\zeta_{t,j}=0$ for $t\ge s+1$ and $j\ge s+1$,
\item if $\zeta_{t,j}\ne 0$ with $t\le s$ and $j\ge s+1$, then $\mathcal{C}_t=\mathbb{F}_q^m$,
\item if $\zeta_{t,j}\ne 0$ with $t\ge s+1$ and $j\le s$, then $\mathcal{C}_j=\mathbb{F}_q^m$,
\item if $\zeta_{t,j}\ne 0$ with $t\le s$ and $j\le s$, then $\mathcal{C}_t^{\perp_E} \subseteq \mathcal{C}_j$.
\end{enumerate}
\end{theorem}

Finally, we state the conditions for dual containment of MPCs in the Hermitian case.

\begin{theorem}
\label{HermitianDualContMP}
Let $\mathcal{C}=\left[\mathcal{C}_1,   \ldots,  \mathcal{C}_s\right] \cdot A$,  where $A$ is of rank $s$ with entries in $\mathbb{F}_{q^2}$ and the codes $\mathcal{C}_i$ are also over $\mathbb{F}_{q^2}$. Consider an $h \times h$ invertible matrix $B$ with entries in $\mathbb{F}_{q^2}$ such that $B\{1,\ldots,s\}=A$. Write $\left(B^{q}B^{\tt t}\right)^{-1}=\left(\zeta_{t,j}\right)_{1 \leq t, j \leq h}$. Then, $\mathcal{C}$ is Hermitian dual-containing if and only if  the following conditions hold.
\begin{enumerate}
\item $\zeta_{t,j}=0$ for $t\ge s+1$ and $j\ge s+1$,
\item if $\zeta_{t,j}\ne 0$ with $t\le s$ and $j\ge s+1$, then $\mathcal{C}_t=\mathbb{F}_{q^2}^m$,
\item if $\zeta_{t,j}\ne 0$ with $t\ge s+1$ and $j\le s$, then $\mathcal{C}_j=\mathbb{F}_{q^2}^m$,
\item if $\zeta_{t,j}\ne 0$ with $t\le s$ and $j\le s$, then $\mathcal{C}_t^{\perp_H} \subseteq \mathcal{C}_j$.
\end{enumerate}
\end{theorem}

\subsection{The minimum distance of the dual of an MPC}

Once the dual of an MPC is known, we can deal with its minimum distance. Note that in order  to do it, we only need to consider the Euclidean case, because Euclidean and Hermitian dual codes are isometric. When  $A$ is an NSC square matrix, this distance can be bounded as the next result shows, see \cite{BlacNor}.

\begin{theorem}
\label{El9}
Let $\mathcal{C}=\left[\mathcal{C}_1,  \ldots,  \mathcal{C}_s\right] \cdot A$, where $A$ is an NSC matrix of size $s\times s$ over $\mathbb{F}_{q}$. Then, the minimum distance of the Euclidean dual of $\mathcal{C}$ satisfies the following equality
\[
d(\mathcal{C}^{\perp_E}) \ge  \min \left\{s d(\mathcal{C}_s^{\perp_E}), (s-1) d(\mathcal{C}_{s-1}^{\perp_E}), \ldots, d(\mathcal{C}_{1}^{\perp_E}) \right\}.
\]
\end{theorem}

A similar result for MPCs given by non-square matrices $A$ is also true.

\begin{corollary}\label{co: DistanceDualMPC}
Keeping the above notation, let  $\mathcal{C}=\left[\mathcal{C}_1,  \ldots, \mathcal{C}_s\right] \cdot A$, where $A$ is an $s \times h$ NSC matrix over $\fq$ with $s \leq h$. Then,

\begin{eqnarray*}
d(\mathcal{C}^{\perp_E}) &\ge&  \min \left\{h, h-1, \ldots, s+1, s d(\mathcal{C}_s^{\perp_E}),\ldots, d(\mathcal{C}_{1}^{\perp_E}) \right\} \\
&=& \min \left\{ s+1, s d(\mathcal{C}_s^{\perp_E}),\ldots, d(\mathcal{C}_{1}^{\perp_E}) \right\}.
\end{eqnarray*}
\end{corollary}
\begin{proof}
Apply Theorem \ref{El9} to the sequence of codes $\mathcal{C}_{1}, \ldots, \mathcal{C}_{s}, \mathcal{O},\ldots, \mathcal{O},$ where $\mathcal{O}$ is the zero vector subspace of $\mathbb{F}_{q}^m$, and use an NSC matrix $B$ of size $h \times h$ such that $B(1,\ldots,s)=A$.
\end{proof}

\subsection{Dual-containing MPCs. The case of square matrices}
\label{La331}

This and the next subsection study specific dual-containing MPCs. We begin dealing with the case when $A$ is a square matrix. We first study Euclidean dual containment given by suitable matrices of size $q$.

In this section and in the following ones we assume
 that $q$ is a power of an odd prime $p$, so that $\mbox{char}(\fq)=p$ and the prime field $\fp$ is a subfield of $\fq$.

A square matrix $M$ of size $s\times s$, $s \leq q$, with entries in $\mathbb{F}_q$,  is called {\em monomial} if it is of the form $M=DP$, where $D$ is a diagonal matrix with nonzero entries in its diagonal, and $P$ is a permutation matrix corresponding to a permutation $\tau$ of $\{1,\dots,s\}$; that this, $P= (p_{i\ell})_{1 \leq i, \ell \leq s}$, $p_{i\ell}=1$ if $\tau(\ell)=i$ and $p_{i\ell}=0$ otherwise.
The following result is due to Cao and Cui \cite[Theorem 3.4 (i)]{Cao1}.

\begin{theorem}\label{monomialTE}
Let $\tau$ be a permutation of $\{1,2,\dots,s\}$ and let $A$ be a non-singular square matrix of size $s \times s$ such that $AA^{\tt t}$ is monomial with respect to the permutation $\tau$. Then the matrix-product code $\mathcal{C}=\left[\mathcal{C}_1,   \ldots,  \mathcal{C}_s\right] \cdot A$ is Euclidean dual-containing if and only if $\mathcal{C}_{\tau(i)}^{\perp_E}\subseteq \mathcal{C}_i$ for all $i$, $1 \leq i \leq s$.
\end{theorem}

As claimed in \cite{Cao2},  in order to apply Theorem \ref{monomialTE}, the main difficulty is to find a matrix $A$ such that $AA^{\tt t}$ is monomial. In that paper, the author proposed the following problem (Problem 5.2 (a)): {\em  to find an NSC matrix $A$ such that $AA^{\tt t}$ is monomial.} In this subsection we give an answer to this problem by showing such a matrix $A$ over $\mathbb{F}_q$ of the largest possible size, which is $q\times q$.

In Subsection \ref{matrixpc} we showed how to obtain NSC matrices. Fix an order on the elements of $\fq$,  so that $\mathbb{F}_q=\{ \alpha_1=0,\alpha_2,\dots,\alpha_q\}$. For  $i \geq 1$, let $\bbw_i$ be the vector obtained by evaluating the monomial $X^{i-1}$ at the points of $\mathbb{F}_q$, that is
$$
\bbw_i=(0^{i-1},\alpha_2^{i-1},\dots,\alpha_q^{i-1})\in\mathbb{F}_q^q.
$$
Let $W$ be the matrix whose rows are $\bbw_1,\dots,\bbw_q$, and  for $1 \leq i \leq q$, let $\mathcal{W}_i$ be the code spanned by the vectors $\bbw_1,\dots,\bbw_i$. It is a Reed-Solomon code of dimension $i$, and thus the matrix $W$ is NSC of size $q\times q$.
The duality properties of the codes $\mathcal{W}_i$ follow from the following  fact.

\begin{lemma}\label{monomial1}
Let $r$ be a nonnegative integer such that $r$ divides $q-1$. Let $\calr$ be the set of zeros of the polynomial $X^{r}-1\in\fp[X]$. Then  $\calr$ is a subset of $\fq$ with $r$ distinct elements. Furthermore, for a nonnegative integer $t$, we have
$$
\mu_{\calr}(t):=\sum_{x\in\calr} x^t=
\begin{cases}
 r\; (\mbox{\rm mod $p$}) & \text{if\, $t\equiv 0\, (\mbox{\rm mod $r$})$,}\\
 0\; (\mbox{\rm mod $p$}) & \text{if\, $t\not\equiv 0\, (\mbox{\rm mod $r$})$.}
\end{cases}
$$
In particular we have
\begin{equation}
\label{ZZ}
\mu(t):= \mu_{\mathbb{F}_q^*} (t) = \sum_{x\in\mathbb{F}_q^*} x^t=
\begin{cases}
 p-1 & \text{if\, $t\equiv 0\, (\mbox{\rm mod $q-1$})$,}\\
 0 & \text{if\, $t\not\equiv 0\, (\mbox{\rm mod $q-1$})$.}
\end{cases}
\end{equation}
\end{lemma}
 
\begin{proof}
$\fq^*$ is the set of zeros of the polynomial  $X^{q-1}-1\in\fp[X]$. Since $r$ divides $q-1$, then  $X^{r}-1$ divides  $X^{q-1}-1$ and $\calr$ is a subset of $\fq$ with $r$ distinct elements. Further, it is easy to see that $\calr$ is a multiplicative subgroup of $\fq^*$, and hence cyclic. Let $\alpha$ be a generator of this group.

If $t\equiv 0\, (\mbox{\rm mod $r$})$ then $x^t=1$ for all $x\in\calr$, so $\mu_{\calr}(t)=r$. If $t\not\equiv 0\, (\mbox{\rm mod $r$})$, we have
$$
\alpha^t \mu_{\calr}(t)=\sum_{x\in\calr} (\alpha x)^t=\mu_{\calr}(t)
$$
as $\alpha x$ runs over all elements of $\calr$. Since $\alpha^t\neq 1$, we conclude that $\mu_{\calr}(t)=0$.

The last statement follows by taking $r=q-1$.
\end{proof}

Let us state the above mentioned duality properties.

\begin{proposition}\label{monomial2}
For $i,j \ge 1$, we have
$$
\langle \bbw_i,\bbw_j \rangle_E=
\begin{cases}
 p-1 & \text{if\, $i+j>2$ and $i+j\equiv 2\, (\mbox{\rm mod $q-1$})$, }\\
  0 & \text{if\, $i+j=2$ or $i+j\not\equiv 2\, (\mbox{\rm mod $q-1$})$. }
\end{cases} 
$$
In particular, if $1\le i,j \le q$, we have $\langle \bbw_i,\bbw_j \rangle_E=-1$ if $i+j=q+1$ or $i+j=2q$, and $\langle \bbw_i,\bbw_j \rangle=0$ otherwise.
\end{proposition}
\begin{proof}
Since $\langle \bbw_i,\bbw_j \rangle_E=0^{i+j-2}+\mu( i+j-2)$,
the result is clearly true for $i=j=1$. Otherwise, according to (\ref{ZZ}),   we have  $\langle \bbw_i,\bbw_j \rangle_E=\mu(i+j-2) =p-1$. 
\end{proof}

As a consequence we obtain the following result.
 
\begin{proposition}\label{monomial3}
For $1\le i< q$, we have $\mathcal{W}_i^{\perp_E}=\mathcal{W}_{q-i}$.
\end{proposition}

Now, for $\ell=1,\dots,q-1$, let $\bba_\ell=\bbw_\ell$, and let $\bba_q=\bbw_q+\lambda\bbw_1$, where $\lambda=(p-1)/2$. For $1 \leq i \leq q$, let $\dot{A}$ be the matrix whose rows are $\bba_1,\dots,\bba_q$, and  $\dot{\mathcal{A}}_i \subseteq \mathbb{F}_q^q$ the code spanned by $\bba_1,\dots,\bba_i$.

\begin{lemma}\label{monomial4}
For $1\le i,j\le q$, we have
$$
\langle \bba_i,\bba_j \rangle_E=
\begin{cases}
 p-1 & \text{if\, $i+j=q+1$, }\\
  0 & \text{if\, $i+j\neq q+1$. }
\end{cases} 
$$
\end{lemma}
 
\begin{proof}
By symmetry it suffices to prove the statement for $i\le j$. Let us consider several cases.\newline
If $i\le j<q$, then $\langle \bba_i,\bba_j\rangle_E=\langle \bbw_i,\bbw_j\rangle_E$.\newline
If $i<j=q$, then $\langle \bba_i,\bba_q\rangle_E=\langle \bbw_i,\bbw_q\rangle_E+\lambda\langle \bbw_i,\bbw_1\rangle_E=\langle \bbw_i, \bbw_q\rangle_E$.\newline
If $i=j=q$, then $\langle \bba_q,\bba_q\rangle_E=\langle \bbw_q,\bbw_q\rangle_E+2\lambda \langle \bbw_1,\bbw_q\rangle_E+\lambda^2\langle \bbw_1,\bbw_1\rangle_E=-1-2\lambda=0$.\newline
Therefore the result follows from Proposition \ref{monomial2}.
\end{proof}

Now, we are ready to give an answer to the above mentioned Problem 5.2 (a) in \cite{Cao2}.
\begin{proposition}\label{monomial5}
The before defined matrix $\dot{A}$ is NSC of size $q\times q$ and such that $\dot{A}\dot{A}^{\tt t}$ is monomial with respect to the permutation $\sigma(i)=q-i+1$. More precisely,  $\dot{A}\dot{A}^{\tt t}$ is the antidiagonal matrix having an entry $p-1$ in each position of its antidiagonal and $0$ elsewhere.
\end{proposition}
\begin{proof}
For all $i=1,\dots,q$, we have $\mathcal{A}_i=\mathcal{W}_i$, hence $\dot{A}$ is NSC.
Our statement concerning $\dot{A}\dot{A}^{\tt t}$ follows from Lemma \ref{monomial4}.
\end{proof}

\begin{example}
For $q=3$, the matrices $W, \dot{A}$ and $\dot{A}\dot{A}^{\tt t}$ are the following ones:
$$
W=\left(\begin{array}{ccc}
1&1&1\\
0&1&2\\
0&1&1
\end{array}\right), \;\;
\dot{A}=\left(\begin{array}{ccc}
1&1&1\\
0&1&2\\
1&2&2
\end{array}\right), \;\;
\dot{A}\dot{A}^{\tt t}=\left(\begin{array}{ccc}
0&0&2\\
0&2&0\\
2&0&0
\end{array}\right).
$$
\end{example}

Considering the before introduced matrix $\dot{A}$ and applying Theorem \ref{monomialTE}, we obtain the following result.

\begin{theorem}\label{monomialTE2}
Let $\mathbb{F}_q$ be a finite field of odd characteristic and keep the above notation. For $i=1, \dots,q$, let $\mathcal{C}_i$ be an  $[m,k_i,d_i]_q$ linear code such that
$\mathcal{C}_{q-i+1}^{\perp_E} \subseteq \mathcal{C}_i$. Then $\mathcal{C}=[\mathcal{C}_1,\ldots,\mathcal{C}_q]\cdot \dot{A}$, is a $[q m, \sum_{1 \leq i \leq q} k_i,\ge d]_q$ Euclidean dual-containing code, where $d \geq \min_{1 \leq i \leq q}\{d_i(q-i+1)\}$.
\end{theorem}

Note that since $\dot{A}\dot{A}^{\tt t}$ is symmetric, it suffices to impose the condition $\mathcal{C}_{q-i+1}^\perp \subseteq \mathcal{C}_i$ for $i=1, 2, \dots, (q+1)/2$. Furthermore, note that our results are confined to the case where $q$ is odd, as the vector $\mathbf{a}_q$ is not well-defined when $q$ is even. Determining the conditions under which these results can be extended to the even case remains an open problem. For instance, when $q=2$, it is straightforward to show that no NSC $2 \times 2$ matrix $A$ exists such that $AA^{\tt t}$ is monomial; this stands in contrast to the case where $q$ is odd, as established in Proposition \ref{monomial5}.\\

Next we treat Hermitian dual containment of MPCs over a field $\mathbb{F}_{q^2}$ of odd characteristic $p$. Recall that the conjugate transpose of a matrix $M$ with entries in $\mathbb{F}_{q^2}$ is denoted $M^{\dagger}$. As before, we recall the following result which can be found in \cite[Theorem 1]{Cao3}; it can also be deduced from Theorem \ref{HermitianDualContMP}.

\begin{theorem}\label{monomialTH}
Let $\mathbb{F}_{q^2}$ be a finite field of odd characteristic. For $i=1,\dots, s$, let $\mathcal{C}_i$ be a linear code over $\mathbb{F}_{q^2}$ such that $\mathcal{C}_{\tau(i)}^{\perp_H}\subseteq \mathcal{C}_i$ for some permutation $\tau$ of $\{1,\dots,s\}$. Let $A$ be a nonsingular matrix of size $s\times s$ with entries in $\mathbb{F}_{q^2}$ and such that $AA^{\dagger}$ is monomial with respect to the permutation $\tau$.
Then $\mathcal{C}=[\mathcal{C}_1,\ldots,\mathcal{C}_s] \cdot A$ is a Hermitian dual-containing code over $\mathbb{F}_{q^2}$.
\end{theorem}

The main difficulty in applying Theorem \ref{monomialTH} is finding a matrix $A$ over $\mathbb{F}_{q^2}$ such that $AA^{\dagger}$ is a monomial matrix. In \cite{Cao2},  the author proposed the following problem (Problem 5.2 (b)): {\em  to  find an NSC matrix $A$ such that $AA^{\dagger}$ is monomial}. Next, we give an answer to this problem showing such a matrix of the largest possible size which is $q^2\times q^2$.

Replacing $q$ with $q^2$, let us consider the previously defined vectors $\bbw_i$ and $\bba_i$. More explicitly, for $i\ge 1$,  $\bbw_i$  is the evaluation of the monomial $X^{i-1}$ at all points in $\mathbb{F}_{q^2}$; for $\ell=1,\dots,q^2-1$, $\bba_\ell=\bbw_\ell$; and  $\bba_{q^2}=\bbw_{q^2}+\lambda\bbw_1$, where $\lambda=(p-1)/2$. A simple computation shows that we have  $\bba_\ell^q=\bbw_{q(\ell-1)+1}$,  for  $\ell=2,\dots,q^2-1$. Furthermore, since $\bba_1$ and $\bba_{q^2}$ have all their coordinates in the prime field $\mathbb{F}_p$, it holds that $\bba_1^q=\bba_1=\bbw_1$ and $\bba_{q^2}^q=\bba_{q^2}=\bbw_{q^2}+\lambda\bbw_1$. As a consequence, from Proposition \ref{monomial2} we get the following result:

\begin{lemma}\label{monomial6}
With the above notation, for $1\le i,j\le q^2$, we have the following equality:
$$
\langle \bba_i,\bba_j \rangle_H=
\begin{cases}
 p-1 & \text{if\, $i+j\neq 2$ and $i+j\neq 2q^2$ and $q(j-1)\equiv q^2-i\, (\mbox{\rm mod $q^2-1$})$, }\\
  0 & \text{if\, $i+j=2$ or $i+j=2q^2$ or $q(j-1)\not\equiv q^2-i\, (\mbox{\rm mod $q^2-1$})$. }
\end{cases}
$$
\end{lemma}

Consider now the matrix $\ddot{A}$ with entries in $\mathbb{F}_{q^2}$ whose rows are  the vectors $\bba_1,\dots,\bba_{q^2}$.

\begin{proposition}\label{monomial7}
The matrix $-\ddot{A}\ddot{A}^\dagger$ is a permutation matrix. Consequently, $\ddot{A}\ddot{A}^\dagger$ is a monomial matrix with respect to the permutation $\sigma$ of $\{1,\dots,q^2\}$ corresponding to the matrix $-\ddot{A}\ddot{A}^\dagger$.
\end{proposition}
\begin{proof}
Since $\langle \mathbf{a}_i, \mathbf{a}_j \rangle_H \in \{0, p-1\}$, it suffices to prove the first statement. Let us show that for any $i$ in the range $1 \le i \le q^2$, there exists a unique $j = j(i)$, where $1 \le j(i) \le q^2$, such that $\langle \mathbf{a}_i, \mathbf{a}_{j(i)} \rangle_H \neq 0$. Furthermore, we shall show that distinct values of $i$ result in distinct values of $j(i)$.

For $i=1$, it follows from Lemma \ref{monomial6} that $\langle \mathbf{a}_1, \mathbf{a}_{q^2} \rangle_H = p-1$ and $\langle \mathbf{a}_1, \mathbf{a}_j \rangle_H = 0$ for all $1 \le j < q^2$; hence, $j(1) = q^2$. Similarly, $\langle \mathbf{a}_{q^2}, \mathbf{a}_1 \rangle_H = p-1$ and $\langle \mathbf{a}_{q^2}, \mathbf{a}_j \rangle_H = 0$ for $1 < j \le q^2$, which implies that $j(q^2) = 1$.

Now, let us consider the case $1 < i < q^2$. First, note that in this case, we have $\langle \mathbf{a}_i, \mathbf{a}_1 \rangle_H = \langle \mathbf{a}_1, \mathbf{a}_i \rangle_H^q = 0$ and $\langle \mathbf{a}_i, \mathbf{a}_{q^2} \rangle_H = \langle \mathbf{a}_{q^2}, \mathbf{a}_i \rangle_H^q = 0$, as shown above. This implies that $j(i) \neq 1, q^2$. Thus, let $i, j$ be such that $1 < i, j < q^2$. These constraints imply $2 < i+j < 2q^2$ and $1 \le q^2-i \le q^2-2$. Under these restrictions, Lemma \ref{monomial6} states that $\langle \mathbf{a}_i, \mathbf{a}_j \rangle_H \neq 0$ if and only if $$q(j-1) \equiv q^2-i \pmod{q^2-1}.$$ Since $\gcd(q, q^2-1) = 1$, $q$ is a unit in $\mathbb{Z}/(q^2-1)\mathbb{Z}$, and therefore the map $\phi: \mathbb{Z}/(q^2-1)\mathbb{Z} \to \mathbb{Z}/(q^2-1)\mathbb{Z}$ defined by $\phi(x) = qx$ induces a bijection of $\{1, 2, \dots, q^2-2\}$ onto itself (since $\phi(x) \equiv 0$ only if $x \equiv 0$). Consequently, for a given $i$ where $1 < i < q^2$, there exists a unique $j$ in the range $1 < j < q^2$ satisfying $q(j-1) \equiv q^2-i \pmod{q^2-1}$; that is, satisfying $\langle \mathbf{a}_i, \mathbf{a}_j \rangle_H \neq 0$. Furthermore, because $\phi$ is a bijection, distinct values of $i$ lead to distinct values of $j$. Thus, we have $j(i) = j$ for these values, and the proof is complete.
\end{proof}
 
Note that the proof of Proposition \ref{monomial7} also provides an explicit expression for the permutation $\sigma$ corresponding to the permutation matrix $-\ddot{A}\ddot{A}^\dagger$, as follows: $\sigma(1)=q^2$, $\sigma(q^2)=1$, and for $1 < i < q^2$,
$$
\sigma(i) = (q^2 - i)q^{-1} \pmod{q^2 - 1}+1 \in \{1, \ldots, q^2\},
$$
where the inversion is taken in $\mathbb{Z}/(q^2-1)\mathbb{Z}$.

\begin{example}
 
For $q=3$, either by performing the product $\ddot{A}\ddot{A}^\dagger$, or from the explicit expression of $\sigma$  we have just presented, we obtain
 
$$
\ddot{A}\ddot{A}^\dagger=\left(\begin{array}{ccccccccc}
0&0&0&0&0&0&0&0&2\\
0&0&0&0&0&2&0&0&0\\
0&0&2&0&0&0&0&0&0\\
0&0&0&0&0&0&0&2&0\\
0&0&0&0&2&0&0&0&0\\
0&2&0&0&0&0&0&0&0\\
0&0&0&0&0&0&2&0&0\\
0&0&0&2&0&0&0&0&0\\
2&0&0&0&0&0&0&0&0
\end{array}\right).
$$
\end{example}

Recall that $\ddot{A}$ is an NSC matrix, as noted above. Thus, considering this matrix, we
can apply Theorem \ref{monomialTH}, obtaining the following result.

\begin{theorem}\label{monomialTH2}
Let $\mathbb{F}_{q^2}$ a finite field of odd characteristic and keep the above notation.  For $i=1, \ldots, q^2$, let $\mathcal{C}_i$ be an $[m,k_i,d_i]_{q^2}$ linear code such that $\mathcal{C}_{\sigma(i)}^{\perp_H}\subseteq \mathcal{C}_i$, where $\sigma$ is the permutation given in Proposition \ref{monomial7}. Then,   $\mathcal{C}=[\mathcal{C}_1,\ldots,\mathcal{C}_q] \cdot \ddot{A}$, is a $[q^2 m, \sum_{1 \leq i \leq q^2} k_i, d]_{q^2}$ Hermitian dual-containing code where
$$
d \geq \min_{1\leq i\leq q^2}\{d_i(q^2+1-i)\}.
$$
\end{theorem}

Note that since $\ddot{A}\ddot{A}^\dagger$ is Hermitian, it suffices to impose the condition $\mathcal{C}_{\sigma(i)}^{\perp_H}\subseteq \mathcal{C}_i$ for $i=1,\dots,(q^2+1)/2$.

\subsection{Dual-containing MPCs. The general case}
\label{La332}

In this subsection, we  take advantage of the results stated in Subsection \ref{La31} to construct Euclidean and Hermitian dual-containing MPCs, by using a matrix $A$, that is not necessarily square, of full rank and size $s \times h$, with $s\le h$. These results can be regarded as extensions of those in Subsection \ref{La331}.  We start with the Euclidean case.

\begin{theorem}\label{te: ConditionsEuclideanDualContaining}
Let $\mathbb{F}_{q}$ be  a finite field of odd characteristic.  Let $h$ and $s$ be two positive  integers such that  $ s\le h<2s$ and 
\begin{itemize}
\item $h-1$ divides $q-1$, and
\item $1-h$ is a square in $\fq$.
\end{itemize}
Also, let $\{\mathcal{C}_j\}_{j=1}^h$ be a family of $q$-ary linear codes satisfying the following conditions:
\begin{itemize}
\item  the parameters of $\mathcal{C}_i$, $1 \leq i \leq s$, are $[m,k_i,d_i]_{q}$ and
\item the following sequence of inclusions and equalities holds:
$$
\mathcal{C}_{h-s+1} \supseteq \mathcal{C}_{h-s+2} \supseteq \cdots \supseteq \mathcal{C}_s \supseteq \mathcal{C}_s^{\perp_E} \mbox{ \;and\; }
\mathcal{C}_1 = \cdots = \mathcal{C}_{h-s} = \mathbb{F}_q^m \mbox{ \;if\; } s<h.
$$
\end{itemize}
Then, there exists an NSC matrix $\dot{A}$ over $\fq$ of size $s \times h$,  such that the matrix-product code $\mathcal{C} = \left[\mathcal{C}_1, \  \ldots, \ \mathcal{C}_s\right] \cdot \dot{A}$ is  Euclidean dual-containing and has parameters $[hm,\sum_{i=1}^s k_i, d]_q$, where
$$
d =  \min \{ hd(\calc_1),\dots, (h+1-s) d(\calc_s)\}.
$$
\end{theorem}
 
\begin{proof}
Let $\calr$ be the set of zeros of the polynomial $X^h-X \in \mathbb{F}_q[X]$.
Since $h-1$ divides $q-1$, then by Lemma \ref{monomial1}, $\calr$ is a subset of $\fq$ with $h$ distinct elements, $\calr=\{\alpha_1=0,\alpha_2,\dots,\alpha_h\}$. Let $\bbw_j$ be the vector obtained by evaluating the monomial $X^{j-1}$ at the points in $\calr$, $j=1,2,\dots,h$. Then the codes $\calr_j$ spanned by $\bbw_1,\dots,\bbw_j$ are Reed-Solomon for all $j$. Since $1-h$ is a square in $\fq$, consider $\beta\in\fq$ such that $\beta^2=1-h$, and define $\bba_1=(\beta,1,\dots,1)\in\fq^h$, $\bba_t=\bbw_j$ for $t=2,\dots,h-1$, and $\bba_h=\bbw_h+\lambda\bba_1$, where $2\lambda+1=0$. Then, for all $j=1,\dots,h$, the code spanned by $\bba_1,\dots,\bba_j$ is the twisted code obtained after multiplying coordinatewise the vectors in $\calr_j$ by
$(\beta,1,\dots,1)$. Then both codes are isometric and thus the matrix $B$ whose rows are the vectors $\bba_1,\dots,\bba_h$ is  NSC.
Using Lemma \ref{monomial1}, a straightforward computation shows that $\langle \bba_i,\bba_j\rangle=h-1$ if $i+j=h+1$ and  $\langle \bba_i,\bba_j\rangle=0$ else. Thus we have
\[
BB^{\tt t}=\begin{pmatrix}
0  & \cdots & 0 & h-1\\
0  & \cdots & h-1 & 0\\
\vdots &   \scalebox{-1}[1]{$\ddots$} & \vdots & \vdots\\
h-1  & \cdots & 0 & 0
\end{pmatrix}.
\]
Let $\dot{A}$ be the matrix formed by the first $s$ rows of $B$, $\dot{A}=B\{1,2,\dots,s\}$.
To prove our statement,  it suffices to show that the conditions given in Theorem~\ref{EuclideanDualContMPC} are satisfied.
Note that $h\neq 1$, and that $(BB^{\tt t})^{-1}$ has nonzero entries only at the positions $(i,j)$ with $i+j=h+1$. Condition (1) of Theorem~\ref{EuclideanDualContMPC} is a consequence of the inequality $h<2s$. For proving Condition (2) note that if $i+j=h+1$ and $j\ge s+1$, then $i\le h-s$, hence $\calc_i=\fq^m$. Similarly for Condition (3). With respect to Condition (4), if $i\le s$ and $j\le s$ then, from the inclusion chain
$$
\mathcal{C}_{h-s+1} \supseteq \mathcal{C}_{h-s+2} \supseteq \cdots \supseteq \mathcal{C}_s \supseteq \mathcal{C}_s^{\perp_E} \mbox{ \;and\; }
\mathcal{C}_1 = \cdots = \mathcal{C}_{h-s} = \mathbb{F}_q^m \mbox{ \;if\; } s<h,
$$
we have
$\calc_i^{\perp_E}\subseteq \calc_s\subseteq\calc_j$.

Finally, the statement about the minimum distance $d$ holds from the fact that $B$, and therefore $\dot{A}$, are NSC and from Proposition \ref{pro:distance}.
\end{proof}

\begin{remark}
{\rm
Note that Theorem \ref{te: ConditionsEuclideanDualContaining} extends Theorem \ref{monomialTE2}. However we stated and proved Theorem \ref{monomialTE2} for the sake of simplicity and because  when the matrix $A$ linked to a matrix-product code is square, we  give a direct consequence of our solution to Problem 5.2 (a) in \cite{Cao2}.
}
\end{remark}

Still in the Euclidean case, we next state a similar result.

\begin{theorem}\label{te: ConditionsEuclideanDualContaining_1}
Let $\mathbb{F}_{q}$ be  a finite field of odd characteristic.
Let $h$ and $s$ be two positive  integers such that
$h$ divides $q-1$ and $ s\le h<2s$.
For $i=1, \ldots, s$, let $\mathcal{C}_i$ be an $[m,k_i,d_i]_{q}$ linear code such that
$$
\mathcal{C}_{h-s+1} \supseteq \mathcal{C}_{h-s+2} \supseteq \cdots \supseteq \mathcal{C}_s \supseteq \mathcal{C}_s^{\perp_E} \mbox{ \;and\; }
\mathcal{C}_1 = \cdots = \mathcal{C}_{h-s} = \mathbb{F}_q^m \mbox{ \;if\; } s<h.
$$
Then there exists an NSC matrix $\dot{A}$ over $\fq$ of size $s \times h$,  such that the matrix-product code $\mathcal{C} = \left[\mathcal{C}_1, \  \ldots, \ \mathcal{C}_s\right] \cdot \dot{A}$ is  Euclidean dual-containing with parameters $[hm,\sum_{i=1}^s k_i, d]_q$, where
$$
d =  \min \{ hd(\calc_1),\dots, (h+1-s) d(\calc_s)\}.
$$
\end{theorem}
\begin{proof}
The proof is quite similar to that of Theorem \ref{te: ConditionsEuclideanDualContaining}. Let $\calr$ be the set of zeros of the polynomial $X^h -1 \in \mathbb{F}_q[X]$. Let $\bba_j$ be the vector obtained by evaluating the monomial $X^{j-1}$ at the points in $\calr$, $j=1,2,\dots,h$. Let $B$ be the matrix whose rows are $\bba_1,\dots,\bba_h$, and $\dot{A}$ the matrix whose rows are $\bba_1,\dots,\bba_s$,  $\dot{A}=B\{1,\dots,s\}$. Clearly, both $\dot{A}$ and $B$ are NSC. From Lemma \ref{monomial1}, we have $\langle \bba_i,\bba_j \rangle=h$ if $i+j=2$ or $i+j=h+2$; and $\langle \bba_i,\bba_j \rangle=0$ else. Thus
$$
BB^{\tt t}=\begin{pmatrix}
h & 0 &  \cdots & 0 & 0\\
0 & 0 &  \cdots & 0 & h\\
0 & 0 & \cdots &  h & 0\\
\vdots& \vdots & \scalebox{-1}[1]{$\ddots$} & \vdots & \vdots\\
0 & h & \cdots & 0& 0
\end{pmatrix}.
$$
Note that $(BB^{\tt t})^{-1}$ has nonzero entries at the same positions as  $BB^{\tt t}$.
Reasoning as in the proof of Theorem \ref{te: ConditionsEuclideanDualContaining}, we deduce that  the conditions of Theorem \ref{EuclideanDualContMPC} are satisfied and hence the result is proved.
\end{proof}

To finish this section we study Hermitian dual-containing MPCs over a field $\mathbb{F}_{q^2}$ in the case $h=q^2$ and $q$ odd. Let $\bba_1,\dots,\bba_{q^2}$ be those vectors with coordinates in $\mathbb{F}_{q^2}$ introduced after Theorem \ref{monomialTH}. According to Lemma \ref{monomial6}, $\langle \bba_i,\bba_j \rangle_H$ equals either $-1$ or $0$. This fact, together with the expression of $\langle \bba_i,\bba_j \rangle_H$ given in that lemma, shows that $\langle \bba_i,\bba_j \rangle_H=\langle \bba_j,\bba_i \rangle_H$.

Let $\sigma$ be a permutation of $\{1,\dots,q^2\}$, denote by $B_{\sigma}$ the $q^2\times q^2$ matrix whose $i$th row is $\bba_{\sigma(i)}$.

\begin{theorem}\label{te: ConditionsHermitianDualContaining}
Let $\mathbb{F}_{q^2}$ be  a finite field of odd characteristic. Let $s$ be an  integer such that
$\frac{q^2+q}{2}<s\le q^2$.
For $i=1, \ldots, s$, let $\mathcal{C}_i$ be an $[m,k_i,d_i]_{q^2}$ linear code such that
$$
\mathcal{C}_{q^2-s+1} \supseteq \mathcal{C}_{q^2-s+2} \supseteq \cdots \supseteq \mathcal{C}_s \supseteq \mathcal{C}_s^{\perp_H} \mbox{ \;and\; }
\mathcal{C}_1 = \cdots = \mathcal{C}_{q^2-s} = \mathbb{F}_{q^2}^m \mbox{ \;if\; } s<q^2.
$$
Then, there exists a full rank matrix $\ddot{A}$ over $\mathbb{F}_{q^2}$ of size $s \times q^2$,  such that the matrix-product code $\mathcal{C} = \left[\mathcal{C}_1,   \ldots,  \mathcal{C}_s\right] \cdot \ddot{A}$ is  Hermitian dual-containing, with parameters  $[q^2m,\sum_{i=1}^s k_i,  d]_{q^2}$,  where
$$
d =  \min \{ d(\calc_1)d(\mathcal{A}_1),\dots, d(\calc_s)d(\mathcal{A}_s)\},
$$
and $\cala_i$ is the code spanned by the first $i$ rows of $\ddot{A}$, $i=1,\dots,s$.
\end{theorem}
\begin{proof}
Let $\bba_1,\dots,\bba_{q^2}$ be as defined above. In the proof of Proposition \ref{monomial7}, we proved that  for any $i$, $1\le i\le q^2$, there exists an unique $j=j(i)$, $1\le j\le q^2$, such that $\langle \bba_i,\bba_j \rangle_H=-1$; and that different values of $i$ lead to different values of $j$. Furthermore, a simple computation using Lemma \ref{monomial6}, shows that $\langle \bba_i,\bba_i \rangle_H=-1$ exactly for $q$ values of $i$, namely $i=q+r(q-1)$, $r=0,1,\dots,q-1$. Thus, there exists a reordering of the $\bba_i$'s
 given by a permutation $\sigma$ of $\{1, \ldots, q^2\}$, such that
 $$
B_{\sigma}^q B_{\sigma}^{\tt t}=
\begin{pmatrix}
-{\rm I}_q & 0\\
0 & B' \\
\end{pmatrix},
$$
$B_{\sigma}$ being as defined before the statement;
the block ${\rm I}_q$ is the identity matrix of size $q \times q$; and the block  $B'$ the following antidiagonal matrix of size $(q^2-q) \times (q^2-q)$:
$$
B'=\begin{pmatrix}
0 &  \cdots & 0 & -1\\
0 &  \cdots &  -1 & 0\\
\vdots& \scalebox{-1}[1]{$\ddots$} & \vdots & \vdots\\
-1 & \cdots & 0& 0
\end{pmatrix}.
 $$
It is simple to see that $(B_{\sigma}^q B_{\sigma}^{\tt t})^2={\rm I}_{q^2}$ and thus $B_{\sigma}^q B_{\sigma}^{\tt t}$ is its own inverse.
Let $\ddot{A}$ be the matrix obtained from the first $s$ rows of $B_{\sigma}$, $\ddot{A}=B_{\sigma}\{1,2,\dots,s\}$.
Reasoning as in the proof of Theorems \ref{te: ConditionsEuclideanDualContaining} and \ref{te: ConditionsEuclideanDualContaining_1}, we deduce that  the conditions in Theorem  \ref{HermitianDualContMP} are satisfied and hence the result is proved. The estimate on the minimum distance $d$ comes from Proposition \ref{pro:distance}.
\end{proof}

Note that, unlike the results stated in the Euclidean case, in Theorem \ref{te: ConditionsHermitianDualContaining} we cannot ensure that the matrix $\ddot{A}$ is NSC when $h<q^2$.
Therefore this matrix does not give an answer to Cao's Problem 5.2 (a) in \cite{Cao2}. Recall also that our results are confined to the case where $q$ is odd. Determining under what conditions they can be translated to the even case seems to be more subtle and remains as an open problem.

\section{Quantum $(r,\delta)$-LRCs from MPCs}
\label{La42}

In this last section, with the help of results developed  in the previous ones, we present some families of quantum $(r,\delta)$-LRCs arising from dual-containing MPCs  based on the three recovery strategies described in Section \ref{La4}. Both the Euclidean and Hermitian dualities will be considered. Exact parameters and localities of  optimal pure quantum codes obtained among them are also given. Throughout this section, $\mathbb{F}_{q}$ will be a field of odd characteristic.

\subsection{Quantum LRCs from Euclidean dual-containing MPCs}
\label{QEuclidean}

We first consider the Euclidean case for dual-containing MPCs.
 
\begin{theorem}\label{te: MainEuclideanCase}
Let $\mathbb{F}_{q}$ be  a finite field of odd characteristic. Let $h, m \leq q$, and $ k=k_1, k_2, \ldots, k_h$ be positive integers such that:
\begin{itemize}
\item $h-1$ divides $q-1$, and  $1-h$ is a square in $\fq$,
\item $m > k=k_1\ge k_2\ge \cdots \ge k_h > m/2$,  and $h > m-k_h$. 
\end{itemize}
 
Then, there exists a pure quantum $(r,\delta)$-LRC, $\calq(\mathcal{C})$, coming from a matrix-product code $\mathcal{C}$, with parameters
$$[[hm,2\sum_{i=1}^h k_i - hm,\ge d]]_q, \mbox{ where } d= \min_{1\le i\le h} \{(m-k_i+1) (h-i+1)\},$$ and locality $(r=k, \delta=m-k+1)$.
\end{theorem}

\begin{proof}
Let $\dot{A}$ be the $h\times h$ matrix with entries in $\mathbb{F}_q$ constructed in Theorem \ref{te: ConditionsEuclideanDualContaining}, where $h=s$ (or $\dot{A}$ the one given in Theorem \ref{monomialTE2}, if $h=q$).
Let  $\calc_1$ be the Reed-Solomon code $\mathrm{RS}(m,k_1)$  and $\calc_t$ the codes $\mathrm{RS}(m,k_t)$, for $ t=2,\ldots, h$. By Proposition \ref{recoveringseparatelyrdelta}, considering  $\calc_1$ as the code $\cald$ in that proposition and since $ \calc_1 \supseteq \calc_2\supseteq \cdots \supseteq \calc_h$, one gets that the code $\mathcal{C}=[\mathcal{C}_1,\ldots,\mathcal{C}_h] \cdot \dot{A}$ is an $(r,\delta)$-LRC with locality $(k,m-k+1)$.

We desire to apply jointly Theorem \ref{Te:QLRC} and Theorem \ref{te: ConditionsEuclideanDualContaining} (or Theorem \ref{monomialTE2}), which will prove the result.
To apply Theorem \ref{Te:QLRC}, we only have to check that $\mathcal{C}^{\perp_E} \subseteq \mathcal{C}$. Indeed, since  $k_ i> m/2$ for every $1 \leq i \leq h$, $\mathcal{C}_i \supseteq \mathcal{C}_i^{\perp_E}$, and we have that
\[
\mathcal{C}_1^{\perp_E}\subseteq \mathcal{C}_2^{\perp_E} \subseteq \cdots  \subseteq
\mathcal{C}_s^{\perp_E}\subseteq \mathcal{C}_s \subseteq \mathcal{C}_{s-1} \subseteq \cdots \subseteq \mathcal{C}_1,
\]
then, the hypotheses of Theorem \ref{te: ConditionsEuclideanDualContaining} (or Theorem \ref{monomialTE2} if $h=q$) hold. Therefore, $\mathcal{C}$ provides a quantum error correcting code (QECC) with parameters  $[[hm,2\sum_{i=1}^h k_i - hm,\ge d]]_q$.
 
The equality
$$
\mathcal{C}^{\perp_E}=[\mathcal{C}_1^{\perp_E}, \ldots, \mathcal{C}_h^{\perp_E}] \cdot (\dot{A}^{-1})^{\tt t}
$$
proved in \cite[Theorem 6.2]{BlacNor}, together with Corollary \ref{co: DistanceDualMPC}, show 
 
\begin{equation}
\label{dualCC}
d(\mathcal{C}^{\perp_E})\geq \min\{h+1, h (k_h +1), \ldots, k_1+1 \}.
\end{equation}
 
If $\min\{h+1, h (k_h +1), \ldots, k_1+1 \} =  \min\{h (k_h +1), \ldots, k_1+1 \}$, then it is larger than or equal to $k_h +1 $. Otherwise, $\min\{h+1, h (k_h +1), \ldots, k_1+1 \} = h+1$ and $h+1 > m-k_h +1 $ because $h > m-k_h$. 

Finally, the equality giving $d$ in the statement and (\ref{dualCC}) prove the following chain of inequalities:
\[
d(\mathcal{C}) = d \leq d (\mathcal{C}_h) \leq m -k_h +1 < d(\mathcal{C}^{\perp_E}).
\]
Therefore, $\calq(\mathcal{C})$ is pure, which concludes the proof.
\end{proof}

Reasoning as in Theorem \ref{te: MainEuclideanCase}, but applying Theorem \ref{te: ConditionsEuclideanDualContaining_1} instead of Theorem \ref{te: ConditionsEuclideanDualContaining}, that is, assuming that $h$ (instead of $h-1$) divides $q-1$,
we get the following result.

\begin{theorem}
\label{te: MainEuclideanCase2}
Let $\mathbb{F}_{q}$ be  a finite field of odd characteristic. Let $h, m \leq q$, and $ k=k_1, k_2, \ldots, k_h$ be positive integers such that:
\begin{itemize}
\item $h$ divides $q-1$,  $1-h$ is a square in $\fq$,
\item $m > k=k_1\ge k_2\ge \cdots \ge k_h > m/2$, and $h > m-k_h$.
\end{itemize}
 
Then, there exist a pure quantum $(r,\delta)$-LRC, $\calq(\mathcal{C})$, coming from a matrix-product code $\mathcal{C}$, with parameters
$$[[hm,2\sum_{i=1}^h k_i - hm,\ge d]]_q, \mbox{ where }d= \min_{1\le i\le h} \{(m-k_i +1) (h-i+1)\},$$ and locality $(r=k, \delta=m-k+1)$.
\end{theorem}
 
As we said after Theorem \ref{Te:QLRC}, a pure quantum $(r,\delta)$-LRC, $\calq(\calc)$, that meets the bound in Inequality (\ref{eq17}) is optimal. For this condition to be met, it suffices to verify that $\calq(\calc)$ is pure and Inequality (\ref{deltaSingletonEquation}) turns out to be an equality for the associated locally recoverable code $\calc$. Our next two corollaries retain the previous notation and determine optimal pure quantum $(r,\delta)$-LRCs.

\begin{corollary}
\label{El36-3}
Let $\mathbb{F}_{q}$ be  a finite field of odd characteristic. For any positive integers $i,j$ such that $(j-1)/2\le i \le j <q/2$,  there exist an optimal pure quantum $(r,\delta)$-LRC with parameters
$$
[[q^2,2((q-i)(q-1)+(q-j))-q^2,j+1]]_q
$$
and locality $(r=q-i, \delta=i+1)$.
\end{corollary}
 
\begin{proof}
Consider the notation as in Theorem \ref{te: MainEuclideanCase}. Set $h=m=q$ and let $\mathcal{C}_1=\cdots=  \mathcal{C}_{q-1}=\mathrm{RS}(q,q-i)$ and $\mathcal{C}_q=\mathrm{RS}(q,q-j)$. Then, apply that theorem noting that the codes $\mathcal{C}_\ell$, for $1 \leq \ell \leq q$, are like the ones considered in its proof . The dimension of the involved MPC, $\mathcal{C}$, is $(q-i)(q-1)+q-j$ and  since $i+1\ge (j+1)/2$, its minimum distance is $d=j+1$. Now, considering $\mathcal{C}$ as an $(r, \delta)$-LRC,  where $r=q-i$ and $\delta=i+1$, one gets
\[
j+1 +(q-1)(q-i) + (q-j) -i + \left( \left\lceil \frac{(q-1)(q-i) + (q-j)}{q-i} \right\rceil \right) i = q^2 +1.
\]
Therefore,  the obtained quantum $(r,\delta)$-LRC $\mathcal{Q}(\mathcal{C})$ is pure, optimal,  and has the parameters given in the statement.
\end{proof}

The following second corollary, extends Corollary \ref{El36-3} to certain integers $h >0$ which include $h=q$ and can be proved by using the same reasoning.

\begin{corollary}
\label{El36-4}
Let $\mathbb{F}_{q}$ be  a finite field of odd characteristic. Let $h$ be  such that $h-1$ divides $q-1$ and $1-h$ is a square in $\mathbb{F}_q$. For any positive integers $i,j$ such that $(j-1)/2\le i \le j <h/2$ and $h^2 - 2 i (h-1) -2j \geq 0$,  there exists an optimal pure quantum $(r,\delta)$-LRC with parameters
$$
[[h^2,2((h-i)(h-1)+(h-j))-h^2,j+1]]_q
$$
and locality $(h-i,i+1)$.

\end{corollary}

We can also
obtain optimal pure quantum $(r,\delta)$-LRCs from MPCs based on the local recovery procedure given by the own matrix $A$, as explained in Subsection \ref{se:SeveralErasuresMB}.

\begin{theorem}\label{te: EuclideanOtimalQLRC}
Let $\mathbb{F}_{q}$ be  a finite field of odd characteristic. For any positive integer $t<\frac{q-1}{3}$, there exists an optimal pure quantum $(r,\delta)$-LRC  $\mathcal{Q}(\mathcal{C})$ with parameters  $$[[q^2, q^2-2qt-4t, 2(t+1)]]_q$$ and locality $(r=q-t,\delta=t+1)$.
\end{theorem}

\begin{proof}
We are going to apply Theorem \ref{te: ConditionsEuclideanDualContaining}. With the notation in that theorem, set $h=m=q$ and $s=q-t$. Then $h <2s$ because $t< \frac{q}{2}$. Define $u= q- 3t$ and consider the NSC matrix $\dot{A}$ introduced in Theorem \ref{te: ConditionsEuclideanDualContaining} and  the sequence of linear codes
\[
\calc_1= \cdots = \calc_{u} = \mathrm{RS}(q,q) \; \mbox{ and } \; \calc_{u+1}= \cdots = \calc_{q-t}=\mathrm{RS}(q,q-1).
\]
 
These codes satisfy the requirements in that theorem; note that $ t < \frac{q}{2}$ implies $\calc_s = \calc_{q-t} \supseteq \calc_{q-t}^{\perp_E}$. In addition, the code $\dot{\cala}$  defined by the matrix $\dot{A}$ has locality $(r=q-t,\delta=t+1)$ (by taking the extended recovery set of all its coordinates).
Therefore, the  MPC $\calc = [\calc_1,\dots,\calc_{q-t}] \cdot \dot{A}$ is Euclidean dual-containing. It has length $q^2$, dimension $$q(q-t)-(q-t-u) = q^2 -qt - 2t$$ and its minimum distance is equal to 
\[
\min \{q, \ldots, q-u+1,2(q-u),\ldots, 2(t+1)\} = \min \{q-u+1, 2(t+1)\} = 2(t+1).
\] 
Note that $t < \frac{q-1}{3}$ ensures that the dimension of the QLRC is positive.

By Corollary \ref{coro12}, the locality of $\calc$ is  $(q-t,t+1)$. The inequality $q-t-u<q-t$ shows that
$$
\left\lceil \frac{k}{r}\right\rceil
=\left\lceil \frac{q(q-t)-(q-t-u)}{q-t}\right\rceil=q.
$$
Thus, the parameters of $\calc$ give an equality in Inequality (\ref{deltaSingletonEquation}) and hence the parameters of $\mathcal{Q}(\mathcal{C})$ reach the bound given in (\ref{eq17}). 

By Corollary \ref{co: DistanceDualMPC}, we know that
\[
d(\calc^{\perp_E}) \ge  \min \{ (q-t+1), (q-t)d^{\perp_E}_{q-t}, \ldots, (u+1) d_{u+1}^{\perp_E} \},
\] 
where $d^{\perp_E}_{\ell}$ means the minimum distance of the Euclidean dual of the code $\calc_\ell$, for $u+1 \leq \ell \leq q-t$. Now, \[
d^{\perp_E}_{q-t}= \cdots = d^{\perp_E}_{u+1} = q.
\] 
Since $\alpha q >q$, for $\alpha \in \{q-t, \ldots, u+1\}$,
\[
d(\calc^{\perp_E}) \ge  \min \{ (q-t+1), q \} = q-t +1.
\]
We conclude by proving the purity of $\mathcal{Q}(\mathcal{C})$ and, therefore, its optimality. It follows from the following sequence of inequalities:
\[
d(\mathcal{C}) = 2(t+1) < q-t +1  \leq d(\mathcal{C}^{\perp_E}),
\]
which holds again  because $t < \frac{q-1}{3}$.
\end{proof}

Theorem \ref{te: EuclideanOtimalQLRC}  can be extended  to provide optimal pure quantum $(r,\delta)$-RLCs of length $m h$, where $ m, h \leq q$ are positive integers.

\begin{theorem}
\label{EEl41}
Let $\mathbb{F}_{q}$ be  a finite field of odd characteristic. Let $m, h \leq q$ be two positive integers such that $h-1$ divides $q-1$ and $1-h$ is a square in $\mathbb{F}_q$. For any positive integer $t < \min\{\frac{h-1}{3}, \frac{mh}{2m+h}\}$,  there exists an optimal pure quantum $(r,\delta)$-LRC  $\mathcal{Q}(\mathcal{C})$ with parameters  $$[[mh, mh -mt-4t, 2(t+1)]]_q$$ and locality $(r=h-t,\delta=t+1)$. 
\end{theorem}

\begin{proof}
The proof is similar to that in the proof of Theorem \ref{te: EuclideanOtimalQLRC}. Here, we consider $u = h - 3t$, the codes $\mathcal{C}_1 =  \cdots = \mathcal{C}_{u} = \mathbb{F}_q^m$ and $\mathcal{C}_{u+1}  = \cdots = \mathcal{C}_{h -t}= \mathrm{RS}[m,m-1]$, and the matrix $\dot{A}$ is that defined in Theorem \ref{te: ConditionsEuclideanDualContaining} for $s=h-t$. The restriction $t < \frac{mh}{2m+h}$ ensures that the dimension of $\mathcal{Q}(\mathcal{C})$ is positive. 
\end{proof}

To complete our study, we now deal with quantum LRCs working according the third  recovery procedure explained in Subection \ref{sect4}. Consider two $q$-ary linear  codes $\mathcal{C}_1$ and $\mathcal{C}_2$  of parameters $[m,k_1,d_1]_q$ and $[m,k_2,d_2]_q$, with $m <q$ and such that
$\mathcal{C}_1^{\perp_E} \subseteq \mathcal{C}_2^{\perp_E}\subseteq \mathcal{C}_2\subseteq \mathcal{C}_1$. Assume that $\mathcal{C}_1$ and $\mathcal{C}_2$ admit enlargements $\hat{\mathcal{C}}_1$ and $\hat{\mathcal{C}}_2$ such that $\hat{\mathcal{C}}_1^{\perp_E} \subseteq \hat{\mathcal{C}}_2^{\perp_E}\subseteq \hat{\mathcal{C}}_2\subseteq \hat{\mathcal{C}}_1$. Consider the code $\hat{\mathcal{C}}$ introduced in Proposition \ref{prop:ext1} attached to a matrix $A$ with $h$ columns. Then, reasoning as in Theorem \ref{te: EuclideanOtimalQLRC} and using Proposition \ref{prop:ext2}, one gets the following result.

\begin{proposition}
\label{LAA43}
Let $\mathbb{F}_{q}$ be  a finite field of odd characteristic.
The above code $\hat{\mathcal{C}}$ gives rise to a quantum $(r,\delta)$-LRC, $\calq(\hat{\mathcal{C}})$, with parameters  $$[[mh+s,2(k_1 + k_2)-mh-s, \ge \min\{d_1 h +s, d_2 (h-1) + s-1 \}]]_q.$$ Furthermore,
if the code  $\cala$ of generator matrix $A$ has locality $(r_\cala,\delta_\cala)$ and $\hat{\mathcal{C}}_1$ has locality $(\hat{r}_1,\hat{\delta}_1)$, then  $\calq(\hat{\mathcal{C}})$  has locality $(r,\delta)$, where $r=\max\{r_\cala,\hat{r}_1\}$ and $\delta = \min \{ \delta_\cala, \hat{\delta}_1 \}$.
\end{proposition}

\subsection{Quantum LRCs from Hermitian dual-containing MPCs}
\label{QHermitian}

All the results presented so far on QLRCs have been based on Euclidean dual containment of MPCs. In this subsection, we will see how we can obtain  quantum $(r,\delta)$-LRCs by considering Hermitian dual-containing MPCs.

A generalized Reed-Solomon code $\mathcal{R}$ of length $m$ and dimension $k$, GRS$(m,k)$ \cite{MAC}, is a linear $[m,k]$-code over $\mathbb{F}_q$ with a parity check matrix of the form $A(m,m-k)D(\alpha_1, \ldots, \alpha_m)$, where $A(m,m-k)$ is a Vandermonde matrix as described in (\ref{matrizRS}) and $D(\alpha_1, \ldots, \alpha_m)$ is
an $m\times m$ diagonal matrix such that $\alpha_1, \ldots, \alpha_m \in \mathbb{F}_q\setminus \{0\}$ are the elements in the diagonal. It is known that there exists a choice of values  $\alpha'_1, \ldots, \alpha'_m \in \mathbb{F}_q\setminus \{0\}$ such that $A(m,k)D(\alpha'_1, \ldots, \alpha'_m)$ is a generator matrix of $\mathcal{R}$. Every GRS$(m,k)$ code is MDS. The following result is a part of Theorem 6 in  \cite{HermitianRS}, where the authors give conditions for the existence of quantum GRS codes coming from Hermitian self-orthogonal GRS codes.

\begin{lemma}\label{le. MDSGRSC}
Let $q$ be a prime power and set $m = q^2-a$ for some integer $0 \leq a \leq q-2$. Let $k$ be an integer such that $q^2-q+1 \le k \le q^2-a$. Then, there exists a generalized Reed-Solomon code $\mathrm{GRS}(m,k)$ such that  $\mathrm{GRS}(m,k)\supseteq$ $\mathrm{GRS}(m,k)^{\perp_H}$.
\end{lemma}

This lemma, together with Proposition \ref{recoveringseparatelyrdelta} and Theorem \ref{monomialTH2}, leads to the following result. Recall that $\ddot{A}$ is the $q^2\times q^2$ NSC matrix over $\mathbb{F}_{q^2}$  defined before Proposition \ref{monomial7}.

\begin{theorem}\label{te: MainHermitianCase}
Let $\mathbb{F}_{q}$ be  a finite field of odd characteristic.  Let $a$ be a nonnegative integer such that $0 \leq a \leq q-2$. Set $m=q^2-a$ and consider positive integers $k_i$ such that
\[
q^2-a \ge k_1=k \ge k_2 \ge k_3 \ge \cdots \ge k_{q^2} > q^2-q+1.
\]
Let $\calc_i$ be codes over $\mathbb{F}_{q^2}$ with parameters  $[m,k_i,d_i]_{q^2}$, for $i=1,\dots,q^2$, as described in Lemma \ref{le. MDSGRSC}. Then, the matrix-product code  $\calc = [\calc_1,\dots,\calc_{q^2}] \cdot \ddot{A}$ gives a quantum $(r,\delta)$-LRC, $\calq(\calc)$, with
parameters $[[q^2 m, 2\sum_{i=1}^{q^2} k_i -q^2 m,\ge d]]_q$, where $d = \min_{1 \leq i \leq q^2} \{d_i(q^2-i+1)\}$, and  locality  $(k,m-k+1)$.
\end{theorem}
\begin{proof}
Use Theorem \ref{monomialTH2}, with $m=q^2 - a$,  together  with codes of suitable dimensions as introduced  in the statement of the theorem. Then Proposition \ref{pro:distance} and  arguments as  in the proof of Theorem \ref{te: MainEuclideanCase} prove the result.
It is convenient to clarify that the matrix used in Theorem \ref{monomialTH2} is NSC but that corresponding to  Hermitian dual code needs not to be NSC. However it is NCS after permutation and then it suffices to bound the minimum distance for the Euclidean dual as we did in Theorem \ref{te: MainEuclideanCase} and use the fact that Euclidean and Hermitian dual are isometric.
\end{proof}

From this theorem we can also obtain optimal pure quantum $(r,\delta)$-LRCs.

\begin{corollary}
\label{El46}
Let $\mathbb{F}_{q}$ be  a finite field of odd characteristic.
Let $a$ and $b$ be nonnegative integers such that $a,b < q-1$ and $a \leq b \leq 2a$. Then there exists an optimal pure quantum $(r,\delta)$-LRC,  $\calq(\mathcal{C})$, with parameters
$$[[q^4,2((q^2-a)(q^2-1)+(q^2-b))-q^4,b+1]]_q$$ and   locality $(q^2-a,a+1)$.
\end{corollary}
\begin{proof}
Apply Theorem \ref{te: MainHermitianCase} to the codes $\mathcal{C}_1=\cdots=\mathcal{C}_{q^2-1}=$ GRS$(q^2,q^2-a)$ and $\mathcal{C}_q=$ GRS$(q^2,q^2-b)$.
Note that $\calq(\mathcal{C})$  comes from a linear code with parameters $$[q^4, (q^2-a)(q^2-1)+(q^2-b), b +1]_{q^2},$$ which meet the bound in Inequality  (\ref{deltaSingletonEquation}). Finally, an argument similar to the one used in the proof of Theorem \ref{te: MainEuclideanCase} establishes the sequence of inequalities
\[
d(\mathcal{C}) = b+1 \leq q^2-q+2 < d(\mathcal{C}^{\perp_H}),
\]
where the first inequality holds since $b < q-1$. Consequently, $\calq(\mathcal{C})$ is an optimal pure quantum $(r,\delta)$-LRC, which concludes the proof.
\end{proof}

We finish this subsection with a table, Table \ref{Tabla1},  that summarizes the parameters of the quantum $(r, \delta)$-LRCs described in this section.

\begin{landscape}

{\tiny
\begin{table}
\begin{tabular}{|c|c|c|c|l|}
\hline
Parameters & Locality & Pure/Optimal & Th./Cor. & Restrictions \\\hline\hline
$[[hm,2\sum_{i=1}^h k_i - hm,\ge d= \min_{1\le i\le h} \{(m-k_i+1) (h-i+1)\}]]_q$& $(k, m-k+1)$ & pure & \ref{te: MainEuclideanCase}& $h-1 |q-1$,  $1-h$ square, $h > m-k_h$ and \\
& & & & $m > k=k_1\ge \cdots \ge k_h > m/2$\\\hline
$[[hm,2\sum_{i=1}^h k_i - hm,\ge d= \min_{1\le i\le h} \{(m-k_i+1) (h-i+1)\}]]_q$& $(k, m-k+1)$ & pure & \ref{te: MainEuclideanCase2}& $h |q-1$,  $1-h$ square, $h > m-k_h$ and \\
& & & & $m > k=k_1\ge \cdots \ge k_h > m/2$\\\hline
$[[q^2,2((q-i)(q-1)+(q-j))-q^2,j+1]]_q$ & $(q-i,i+1)$ & optimal pure & \ref{El36-3} & $(j-1)/2\le i \le j <q/2$\\\hline
$[[h^2,2((h-i)(h-1)+(h-j))-h^2,j+1]]_q$ & $(h-i,i+1)$ & optimal pure & \ref{El36-4} & $h-1 | q-1$ and $1-h$ square\\
& & & &  $(j-1)/2\le i \le j <h/2$ and $h^2 - 2 i (h-1) -2j \geq 0$\\\hline
$[[q^2, q^2 - 2qt - 4t, 2(t+1)]]_q$ & $(q-t,t+1)$ & optimal pure & \ref{te: EuclideanOtimalQLRC} & $t<\frac{q-1}{3}$ \\\hline
$[[mh, mh -2mt -4t, 2(t+1)]]_q$ & $(h-t,t+1)$ & optimal pure & \ref{EEl41} & $h-1 | q-1$, $1-h$ square and $t < \min\{\frac{h-1}{3}, \frac{mh}{2m+h}\}$\\\hline
$[[q^2 m, 2\sum_{i=1}^{q^2} k_i -q^2 m,\ge d= \min_{1 \leq i \leq q^2} \{d_i(q^2-i+1)\}]]_q$ & $(k,m-k+1)$ & - & \ref{te: MainHermitianCase} & $0 \leq a \leq q-2$, $m=q^2-a$ and \\
& & & &  $q^2-a \ge k_1=k \ge \cdots \ge k_{q^2} > q^2-q+1$ \\\hline
$[[q^4,2((q^2-a)(q^2-1)+(q^2-b))-q^4,b+1]]_q$ & $(q^2-a,a+1)$ & optimal pure & \ref{El46} & $a,b < q-1$ and $a \leq b \leq 2a$
\\\hline
\end{tabular}
\caption{Parameters and localities of $q$-ary quantum $(r,\delta)$-LRCs, $q$ odd}
\label{Tabla1}
\end{table}}
\end{landscape}

\subsection{Examples}
\label{laultima}

In this final subsection, we provide three tables (Tables \ref{Tabla2}, \ref{Tabla3}, and \ref{Tabla4}) containing the parameters and localities of several optimal pure quantum $(r,\delta)$-locally recoverable codes obtained using our results. We focus on $q$-ary codes for small values of $q$, namely $q=5,7,$ and $9$. As this is an emerging topic, no similar lists currently exist for comparison; consequently, these values may serve as a benchmark for future research. For instance, they may help determine which integers permitted by the bound in Inequality (\ref{eq17}) of Theorem \ref{Te:QLRC} can be effectively realized by an optimal pure quantum $(r,\delta)$-LRC. A considerable number of these values are derived from Theorem \ref{EEl41}, which is based on the results of Subsection \ref{se:SeveralErasuresMB} regarding a novel strategy for the (classical) local recovery of matrix-product codes. In the tables, the column labeled `Th./Cor.' refers to the specific Theorem or Corollary providing the corresponding codes.

{\tiny
\begin{table}[h]
\begin{tabular}{|ccc|ccc|ccc|}
\hline
parameters & Locality & Th./Cor.  & Parameters & Locality & Th./Cor. & Parameters & Locality & Th./Cor. \\\hline
\rule{0mm}{5mm}\!\!
$[[9, 3, 2]]$  & $ (2, 2)$ & \ref{El36-4} & $[[25, 11, 4]]$ & $(4, 2)$   & \ref{EEl41}, \ref{te: EuclideanOtimalQLRC} & $[[625, 523, 4]]$ & $(23, 3)$ & \ref{El46} \\
$[[10, 2, 4]]$ & $(4, 2)$ & \ref{EEl41} & $[[25, 13, 3]]$ & $(4, 2)$    & \ref{El36-3} & $[[625, 525, 3]]$ & $(23, 3)$ & \ref{El46} \\
$[[15, 5, 4]]$ & $(4, 2)$ & \ref{EEl41} & $[[25, 15, 2]]$ & $(4, 2)$    & \ref{El36-3} & $[[625, 573, 3]]$ & $(24, 2)$ & \ref{El46} \\
$[[20, 8, 4]]$ & $(4, 2)$ & \ref{EEl41} & $[[625, 475, 4]]$ & $(22, 4)$ & \ref{El46} & $[[625, 575, 2]]$ & $(24, 2)$   & \ref{El46} \\
$[[25, 5, 3]]$ & $(3, 3)$ & \ref{El36-3} & &&&&\\ \hline
\end{tabular}
\caption{\rule{0mm}{6mm}Parameters and localities of optimal quantum LRCs for $q=5$.}
\label{Tabla2}
\end{table}}

{\tiny
\begin{table}[h]
\begin{tabular}{|ccc|ccc|ccc|}
\hline
Parameters & Locality & Th./Cor.  & Parameters & Locality & Th./Cor. & Parameters & Locality & Th./Cor. \\\hline
\rule{0mm}{5mm}\!\!
$[[7, 1, 4]]$ & $(6, 2)$ & \ref{EEl41} & $[[49, 13, 6]]$ & $(5, 3)$ & \ref{te: EuclideanOtimalQLRC} & $[[2401, 2103, 6]]$ & $(46, 4)$ & \ref{El46} \\
$[[9, 3, 2]]$ & $ (2, 2)$ & \ref{El36-4} & $[[49, 19, 4]]$ & $(5, 3)$ & \ref{El36-3} & $[[2401, 2105, 5]]$ & $(46, 4)$ & \ref{El46} \\
$[[14, 6, 4]]$ & $(6, 2)$ & \ref{EEl41} & $[[49, 21, 3]]$ & $(5, 3)$ & \ref{El36-3} & $[[2401, 2107, 4]]$ & $(46, 4)$ & \ref{El46} \\
$[[16, 8, 2]]$ & $ (3, 2)$ & \ref{El36-4} & $[[49, 31, 4]]$ & $(6, 2)$ & \ref{El36-3},\ref{te: EuclideanOtimalQLRC},\ref{EEl41} & $[[2401, 2201, 5]]$ & $(47, 3)$ & \ref{El46} \\
$[[21, 11, 4]]$ & $(6, 2)$ & \ref{EEl41} & $[[49, 33, 3]]$ & $(6, 2)$ & \ref{El36-3} & $[[2401, 2203, 4]]$ & $(47, 3)$ & \ref{El46} \\
$[[28, 16, 4]]$ & $(6, 2)$ & \ref{EEl41} & $[[49, 35, 2]]$ & $(6, 2)$ & \ref{El36-3} & $[[2401, 2205, 3]]$ & $(47, 3)$ & \ref{El46} \\
$[[35, 21, 4]]$ & $(6, 2)$ & \ref{EEl41} & $[[2401, 1911, 6]]$ & $(44, 6)$ & \ref{El46} & $[[2401, 2301, 3]]$ & $(48, 2)$ & \ref{El46} \\
$[[42, 26, 4]]$ & $(6, 2)$ & \ref{EEl41} & $[[2401, 2007, 6]]$ & $(45, 5)$ & \ref{El46} & $[[2401, 2303, 2]]$ & $(48, 2)$ & \ref{El46} \\
$[[49, 7, 4]]$ & $(4, 4)$ & \ref{El36-3} & $[[2401, 2009, 5]]$ & $(45, 5)$ & \ref{El46} & & & \\
\hline
\end{tabular}
\caption{\rule{0mm}{6mm}Parameters and localities of optimal quantum LRCs for $q=7$.}
\label{Tabla3}
\end{table}}

{\tiny
\begin{table}[h]
\begin{tabular}{|ccc|ccc|ccc|}
\hline
Parameters & Locality & Th./Cor. & Parameters & Locality & Th./Cor. & Parameters & Locality & Th./Cor. \\\hline
\rule{0mm}{5mm}\!\!
$[[9, 3, 2]]$ & $ (2, 2)$ & \ref{El36-4} & $[[45, 23, 4]]$ & $(4, 2)$ & \ref{EEl41} & $[[6561, 5587, 8]]$ & $(75, 7)$ & \ref{El46} \\
$[[9, 3, 4]]$ & $(8, 2)$ & \ref{EEl41} & $[[45, 31, 4]]$ & $(8, 2)$ & \ref{EEl41} & $[[6561, 5589, 7]]$ & $(75, 7)$ & \ref{El46} \\
$[[10, 2, 4]]$ & $(4, 2)$ & \ref{EEl41} & $[[54, 22, 6]]$ & $(7, 3)$ & \ref{EEl41} & $[[6561, 5747, 8]]$ & $(76, 6)$ & \ref{El46} \\
$[[15, 5, 4]]$ & $(4, 2)$ & \ref{EEl41} & $[[54, 38, 4]]$ & $(8, 2)$ & \ref{EEl41} & $[[6561, 5749, 7]]$ & $(76, 6)$ & \ref{El46} \\
$[[18, 2, 6]]$ & $(7, 3)$ & \ref{EEl41} & $[[63, 27, 6]]$ & $(7, 3)$ & \ref{EEl41} & $[[6561, 5751, 6]]$ & $(76, 6)$ & \ref{El46} \\
$[[18, 10, 4]]$ & $(8, 2)$ & \ref{EEl41} & $[[63, 45, 4]]$ & $(8, 2)$ & \ref{EEl41} & $[[6561, 5907, 8]]$ & $(77, 5)$ & \ref{El46} \\
$[[20, 8, 4]]$ & $(4, 2)$ & \ref{EEl41} & $[[72, 32, 6]]$ & $(7, 3)$ & \ref{EEl41} & $[[6561, 5909, 7]]$ & $(77, 5)$ & \ref{El46} \\
$[[25, 5, 3]]$ & $ (3, 3)$ & \ref{El36-4} & $[[72, 52, 4]]$ & $(8, 2)$ & \ref{EEl41} & $[[6561, 5911, 6]]$ & $(77, 5)$ & \ref{El46} \\
$[[25, 11, 4]]$ & $(4, 2)$ & \ref{EEl41} & $[[81, 9, 5]]$ & $(5, 5)$ & \ref{El36-3} & $[[6561, 5913, 5]]$ & $(77, 5)$ & \ref{El46} \\
$[[25, 13, 3]]$ & $ (4, 3)$ & \ref{El36-4} & $[[81, 25, 5]]$ & $(6, 4)$ & \ref{El36-3} & $[[6561, 6069, 7]]$ & $(78, 4)$ & \ref{El46} \\
$[[25, 15, 2]]$ & $ (4, 2)$ & \ref{El36-4} & $[[81, 27, 4]]$ & $(6, 4)$ & \ref{El36-3} & $[[6561, 6071, 6]]$ & $(78, 4)$ & \ref{El46} \\
$[[27, 7, 6]]$ & $(7, 3)$ & \ref{EEl41} & $[[81, 37, 6]]$ & $(7, 3)$ & \ref{te: EuclideanOtimalQLRC},\ref{EEl41} & $[[6561, 6073, 5]]$ & $(78, 4)$ & \ref{El46} \\
$[[27, 17, 4]]$ & $(8, 2)$ & \ref{EEl41} & $[[81, 41, 5]]$ & $(7, 3)$ & \ref{El36-3} & $[[6561, 6075, 4]]$ & $(78, 4)$ & \ref{El46} \\
$[[30, 14, 4]]$ & $(4, 2)$ & \ref{EEl41} & $[[81, 43, 4]]$ & $(7, 3)$ & \ref{El36-3} & $[[6561, 6233, 5]]$ & $(79, 3)$ & \ref{El46} \\
$[[35, 17, 4]]$ & $(4, 2)$ & \ref{EEl41} & $[[81, 45, 3]]$ & $(7, 3)$ & \ref{El36-3} & $[[6561, 6235, 4]]$ & $(79, 3)$ & \ref{El46} \\
$[[36, 12, 6]]$ & $(7, 3)$ & \ref{EEl41} & $[[81, 59, 4]]$ & $(8, 2)$ & \ref{El36-3}, \ref{te: EuclideanOtimalQLRC}, \ref{EEl41} & $[[6561, 6237, 3]]$ & $(79, 3)$ & \ref{El46} \\
$[[36, 24, 4]]$ & $(8, 2)$ & \ref{EEl41} & $[[81, 61, 3]]$ & $(8, 2)$ & \ref{El36-3} & $[[6561, 6397, 3]]$ & $(80, 2)$ & \ref{El46} \\
$[[40, 20, 4]]$ & $(4, 2)$ & \ref{EEl41} & $[[81, 63, 2]]$ & $(8, 2)$ & \ref{El36-3} & $[[6561, 6399, 2]]$ & $(80, 2)$ & \ref{El46} \\
$[[45, 17, 6]]$ & $(7, 3)$ & \ref{EEl41} & $[[6561, 5427, 8]]$ & $(74, 8)$ & \ref{El46} & & & \\
\hline
\end{tabular}
\caption{\rule{0mm}{6mm}Parameters and localities of optimal quantum LRCs for $q=9$.}
\label{Tabla4}
\end{table}}
 
\section*{Acknowledgments}
The authors thank the editor and anonymous reviewers for their helpful comments and feedback, which improved this manuscript.


\begin{thebibliography}{00}

\bibitem{Aly}
S.A. Aly, A. Klappenecker, P.K. Sarvepalli,
On quantum and classical BCH codes,
IEEE Transactions on Information Theory 53 (2007), 1183-1188.

\bibitem{AND}
S.E. Anderson, E. Camps-Moreno, H.H. López, G.L. Matthews, D. Ruano and I. Soprunov,
Relative hulls and quantum codes,
IEEE Transactions on Information Theory 70 (2024), 3190-3201.

\bibitem{Ashi}
A. Ashikhmin, E. Knill,
Nonbinary quantum stabilizer codes,
IEEE Transactions on Information Theory 47 (2001), 3065-3072.

\bibitem{BlacNor}
T. Blackmore, G.H. Norton,
Matrix-product codes over $\fq$,
Applicable Algebra in Engineering, Communication and Computing 12 (2001), 477-500.

\bibitem{CMST}
H. Cai, Y. Miao, M. Schwartz, X. Tang,
On optimal locally repairable codes with superlinear length,
IEEE Transactions on Information Theory 66 (2020), 4853-4868.

\bibitem{CAIFAN}
H. Cai, C. Fan, Y. Miao, M. Schwartz, X. Tang,
Optimal locally repairable codes: an improved bound and constructions,
IEEE Transactions on Information Theory 68 (2022), 5060-5074.

\bibitem{Calder1}
A.R. Calderbank, P.W. Shor,
Good quantum error-correcting codes exist,
Physical Review A 54 (1996), 1098--1105.

\bibitem{Calder2}
A.R. Calderbank, E.M. Rains, P.W. Shor,  N.J.A. Sloane,
Quantum error correction via codes over GF(4),
IEEE Transactions on Information Theory 44 (1998), 1369-1387.

\bibitem{Cao2} M. Cao, Quantum error-correcting codes from matrix-product codes related to quasi-orthogonal and quasy-unitary matrices, Arxiv:2012.15691 (2022).

\bibitem{Cao3} M. Cao, J. Cui, Construction of new quantum codes via Hermitian dual-containing matrix-product codes, Quantum Information Processing 19 (2020), 1-26.

\bibitem{Cao1} M. Cao, J. Cui, New stabilizer codes from the construction of dual-containing matrix product codes, Finite Fields and their Applications 63  (2020), 1017643.

\bibitem{CaoL} M. Cao, K. Zhou, Optimal quantum $(r,\delta)$-locally repairable codes from matrix-product codes, IEEE Transactions on Information Theory 72 (2026) 5572-5591. 

\bibitem{CH}
B. Chen, J. Huang,
A construction of optimal $(r,\delta)$ locally recoverable codes,
IEEE Access 7 (2019), 180349-180353.

\bibitem{CXHF}
B. Chen, S-T. Xia, J. Hao, F-W. Fu,
Constructions of optimal cyclic $(r, \delta)$ locally repairable codes,
IEEE Transactions on Information Theory 64 (2018), 2499-2511.

\bibitem{FAN}
Y. Fan, S. Ling, H. Liu,
Homogeneous weights of matrix-product codes over finite principal ideal rings,
Finite Fields and their Applications 29 (2014), 247-267.

\bibitem{FLL}
Y. Fan, S. Ling, H. Liu,
Matrix-product codes over finite commutative Frobenius rings,
Designs, Codes and Cryptography 71 (2014), 201–227.

\bibitem{FF}
W. Fang, F. Fu,
Optimal cyclic $(r,\delta)$ locally recoverable codes with unbounded length,
Finite Fields and their Applications 63 (2020), 1-14.
 

\bibitem{HE}
C. Galindo, F. Hernando, H. Mart\'{\i}n-Cruz,
Optimal $(r,\delta)$-LRCs from monomial-Cartesian codes and their subfield-subcodes,
Designs Codes and Cryptography 92 (2024), 2549–2586.

\bibitem{QLRC} C. Galindo, F. Hernando, H. Martín-Cruz, R. Matsumoto,
Quantum $(r,\delta)$-locally recoverable codes,
Finite Fields and their Applications 111 (2026), 102785.


\bibitem{HER-HE}
C. Galindo, F. Hernando, H. Mart\'{\i}n-Cruz,
New quantum codes from homothetic-BCH codes,
Computational and Applied Mathematics 45 (2026), 132.

\bibitem{GHM}
C. Galindo, F. Hernando,  C. Munuera,
Locally recoverable $J$-affine variety codes,
Finite Fields and their Applications 64 (2020), 101661,


\bibitem{GHR}
C. Galindo, F. Hernando,  D. Ruano,
New quantum codes from evaluation and matrix-product codes,
Finite Fields and their Applications 36 (2015), 98-120.

\bibitem{GUR2}
L. Golowich, V. Guruswami,
Quantum locally recoverable codes,
arXiv:2311.08653v1, 2025.

\bibitem{GUR}
L. Golowich, V. Guruswami,
Quantum locally recoverable codes,
in Proc. Annu. ACM-SIAM Symp. Discrete Algorithms (SODA),
5512–5522, 2025.

\bibitem{GHSY}
P. Gopalan, C. Huang, H. Simitci, S. Yekhanin,
On the locality of codeword symbols,
IEEE Transactions on Information Theory 58 (2012), 6925-6934.

\bibitem{Gott}
D. Gottesman,
Class of quantum error-correcting codes saturating the quantum Hamming bound,
Physical Review A 54 (1996), 1862-1868.
 

\bibitem{HLR}
F. Hernando, K. Lally, D. Ruano,
Construction and decoding of matrix product codes from nested codes,
Applicable Algebra in Engineering, Communication and Computing 20 (2009), 497-507.

\bibitem{HER}
F. Hernando, D. Ruano,
Decoding of matrix-product codes,
Journal of Algebra and Its Applications 12 (2013), 1250185.

\bibitem{Jin}
L. Jin,
Explicit construction of optimally locally recoverable codes codes of distance 5 and 6 via binary constant weight codes,
IEEE Transactions on Information Theory 65 (2019) 4658-4663.

\bibitem{KKK}
A. Ketkar, A. Klapenecker, S. Kumar, P. Sarpevalli,
Nonbinary tabilizer codes over finite fields,
IEEE Transactions on Information Theory 52 (2006), 4892-4914.
 

\bibitem{KWG}
X. Kong, X. Wang, G. Ge,
New constructions of optimal locally repairable codes with superlinear length,
IEEE Transactions on Information Theory  67 (2021),  6491-6506.

\bibitem{LAG}
G.G. La Guardia,
On the construction of nonbinary quantum BCH codes,
IEEE Transactions on Information Theory  60 (2014),  1528-1535.

\bibitem{HermitianRS} Z. Li, L. Xing, X. Wang, Quantum generalized Reed-Solomon codes: unified framework for quantum maximum-distance-separable codes, Physical Review A 77 (2008).

\bibitem{LIUDIN}
X. Liu et al.,
On new quantum codes from matrix product codes.
Cryptography and Communications 10 (2018), 579–589.

\bibitem{LMC}
J. Liu, S. Mesnager, L. Chen,
New constructions of optimal locally recoverable codes via good polynomials,
IEEE Transactions on Information Theory  64 (2018),  889-899.

\bibitem{LMD}
J. Liu, S. Mesnager, D. Tang,
Constructions of optimal locally recoverable codes via Dickson polynomials,
Designs, Codes and Cryptography 88 (2020),  1759-1780.

\bibitem{LEL}
G. Luo, M.F. Ezerman, S. Ling,
Three new constructions of optimal locally
repairable codes from matrix-product codes,
IEEE Transactions on Information Theory 69 (2023),  75-85.

\bibitem{LXY}
Y. Luo, C. Xing, C. Yuan,
Optimal locally repairable codes of distance 3 and 4 via cyclic codes,
IEEE Transactions on Information Theory, 69 (2023), 75-85.

\bibitem{LUO}
G. Luo, B. Chen, M.F. Ezerman, S. Ling,
Bounds and constructions of quantum locally recoverable codes from quantum CSS codes,
IEEE Transactions on Information Theory, 71 (2025), 1794-1802.


\bibitem{MAC}
F.J. MacWilliams, N.J.A. Sloane,
The theory of error-correcting codes,
Elsevier-North-Holland, Amsterdam, 1977.

\bibitem{Mi}
G. Micheli,
Constructions of locally recoverable codes which are optimal,
IEEE Transactions on Information Theory  66  (2020),  167-175.

\bibitem{chuangnielsen}
M.A. Nielsen, I.L. Chuang,
Quantum computation and quantum information,
Cambridge University Press, 2001.

\bibitem{OS}
F. \"Ozbudak, H. Stichtenoth,
Note on Niederreiter-Xing's propagation rule for linear codes,
Applicable Algebra in Egeneering, Communication and Computing 13 (2002), 53-56.

\bibitem{PKLK}
N. Prakash, G.M. Kamath, V. Lalitha and P.V. Kumar,
Optimal linear codes with a local-error-correction property,
Proceedings of IEEE Int. Symp. Inform. Theory (ISIT-2012), 2776-2780, Boston 2012.

\bibitem{PLK}
N. Prakash, V. Lalitha and P.V. Kumar,
Codes with locality for two erasures,
Proceedings of IEEE Int. Symp. Inform. Theory (ISIT-2014), 1962-1966, Honolulu 2014.

\bibitem{handbook}
V. Ramkumar, M. Vajha, S.B. Balaji, M.N.  Krishnan, B. Sasidharan, and P.V. Kumar, Codes for distributed storage, in Concise encyclopedia of coding theory, 735-761, CRC Press 2021.

\bibitem{SVAV}
C. Salgado, A. Varilly-Alvarado, J.F. Voloch,
Locally recoverable codes on surfaces,
IEEE Transactions on Information Theory 67 (2021), 5765-5777.

\bibitem{ROD}
R. San-Jos\'e,
About the generalized Hamming weights of matrix-product codes,
Computational and Applied Mathematics 44 (2025), 186.

\bibitem{SHAR}
S. Sharma, V. Ramkumar, I. Tamo,
Quantum locally recoverable codes via good polynomials,
IEEE Journal on Selected Areas in Information Theory 6 (2025), 100-110.

\bibitem{SOB}
R. Sobhani,
Matrix-product structure of repeated-root cyclic codes over finite fields,
Finite Fields and their Applications 39 (2016), 216-232.

\bibitem{STE} A.M. Steane,
Simple quantum error correcting codes,
Physical Review A 54 (1996), 4741.

\bibitem{Dual MPC} R. Taki-Eldin,
Galois duality of matrix product codes with applications,
Journal of Computational and Applied Mathematics  474 (2026), 116954.


\bibitem{TB}
I. Tamo, A. Barg,
A family of optimal locally recoverable codes,
IEEE Transactions on Information Theory  60  (2014),  4661-4676.

\bibitem{TPD}
I. Tamo, D.S. Papailiopoulos, A.G. Dimakis,
Optimal locally repairable codes and connections to matroid theory,
IEEE Transactions on Information Theory  62  (2016),  6661-6671.

\bibitem{VAN}
B. van Asch,
Matrix-product codes over finite chain rings,
Applicable Algebra in Egeneering, Communication and Computing  19,  (2008), 39–49.

\bibitem{WZ}
A. Wang and Z. Zhang,
Repair locality with multiple erasure tolerance,
IEEE Transactions on Information Theory 60 (2014), 6979-6987.

\bibitem{Zh}
G. Zhang,
A new construction of optimal $(r,\delta)$ locally recoverable codes,
IEEE Communications Letters 66 (2020), 7333-7340.

\bibitem{ZL}
G. Zhang, H. Liu,
Constructions of optimal codes with hierarchical locality,
IEEE Transactions on Information Theory 66 (2020), 7333-7340.

\bibitem{ZXL}
Z. Zhang, J. Xu, M. Liu:
Constructions of optimal locally repairable codes over small fields,
Scientia Sinica Math.  47 (2017), 1607-1614.

\bibitem{Wang}
L. Wang, X. Zhang, S. Zhu,
Matrix-product constructions for Galois self-orthogonal codes and new quantum codes.
Computational and  Applied  Mathematics 44 (2025), 370.

\bibitem{ZHANG}
W. Zhang, C. Yuan, Y. Luo, N. Li,
Bounds on the size of $(r,\delta)$-locally repairable codes for fixed values $q$ and $d$,
Finite Fields and Their Applications 109 (2026), 102697.

\bibitem{ZHOU}
K. Zhou, M. Cao,
Optimal quantum $(r,\delta)$-locally repairable codes via classical ones,
Journal of Combinatorial Theory, Series A 223 (2026), 106212.

\end{thebibliography}
\end{document}